\documentclass[journal]{IEEEtran}

\ifCLASSOPTIONcompsoc
  \usepackage[nocompress]{cite}
\else
  \usepackage{cite}
\fi

\ifCLASSINFOpdf
\else
\fi
\usepackage{amsmath}
\usepackage{amssymb}
\usepackage{algorithmic}
\usepackage{algorithm}
\usepackage{enumerate}
\usepackage{graphicx}
\usepackage{color}
\usepackage{multirow,tabularx}
\usepackage{enumerate}
\usepackage{times}
\usepackage{booktabs}
\usepackage{subfloat}

\ifCLASSOPTIONcompsoc
  \usepackage[caption=false,font=footnotesize,labelfont=sf,textfont=sf]{subfig}
\else
  \usepackage[caption=false,font=footnotesize]{subfig}
\fi

\usepackage{fixltx2e}
\usepackage{dblfloatfix}
\usepackage{url}
\usepackage{bm}
\usepackage{multirow,tabularx}
\usepackage[subfigure]{tocloft}
\newcolumntype{C}{>{\centering\arraybackslash}X}

\usepackage{amsthm}

\theoremstyle{plain}

\theoremstyle{definition}

\usepackage[breaklinks=true,bookmarks=false]{hyperref}

\renewcommand{\algorithmicrequire}{\textbf{Input:}}
\renewcommand{\algorithmicensure}{\textbf{Output:}}

\begin{document}
\title{Epipolar Focus Spectrum: A Novel Light Field Representation and Application in Dense-view Reconstruction}

\author{Yaning Li,
        Xue Wang,
        Hao Zhu,
        Guoqing Zhou, and
        Qing Wang,~\IEEEmembership{Senior Member,~IEEE}
        %, and        Jingyi Yu%,~\IEEEmembership{Fellow,~IEEE} % <-this % stops a space

\IEEEcompsocitemizethanks{\IEEEcompsocthanksitem 
Y. Li, X. Wang, G. Zhou, Q. Wang (corresponding author) are with the School of Computer Science, Northwestern Polytechnical University, Xi'an 710072, China. E-mail: qwang@nwpu.edu.cn.
\IEEEcompsocthanksitem  H. Zhu is with the School of Electronic Science and Engineering, Nanjing University, Nanjing 210023, China. E-mail: zhuhao\_photo@nju.edu.cn.
%\IEEEcompsocthanksitem J. Yu is with ShanghaiTech University, Shanghai 200031, China. E-mail: jingyi.udel@gmail.com.
\IEEEcompsocthanksitem The work was supported by NSFC under Grant 62031023, Grant 61801396 and Grant 62101242.}% <-this % stops an unwanted space
\thanks{Manuscript received Dec 2021.}}


\markboth{Journal of \LaTeX\ Class Files,~Vol.~14, No.~8, August~2015}%
{Shell \MakeLowercase{\textit{et al.}}: Bare Demo of IEEEtran.cls for Computer Society Journals}



\IEEEtitleabstractindextext{%
\begin{abstract}
Existing light field representations, such as epipolar plane image (EPI) and sub-aperture images, do not consider the structural characteristics across the views, so they usually require additional disparity and spatial structure cues for follow-up tasks. Besides, they have difficulties dealing with occlusions or larger disparity scenes. To this end, this paper proposes a novel Epipolar Focus Spectrum (EFS) representation by rearranging the EPI spectrum. Different from the classical EPI representation where an EPI line corresponds to a specific depth, there is a one-to-one mapping from the EFS line to the view. Accordingly, compared to a sparsely-sampled light field, a densely-sampled one with the same field of view (FoV) leads to a more compact distribution of such linear structures in the double-cone-shaped region with the identical opening angle in its corresponding EFS. Hence the EFS representation is invariant to the scene depth. To demonstrate its effectiveness, we develop a trainable EFS-based pipeline for light field reconstruction, where a dense light field can be reconstructed by compensating the ``missing EFS lines'' given a sparse light field, yielding promising results with cross-view consistency, especially in the presence of severe occlusion and large disparity. Experimental results on both synthetic and real-world datasets demonstrate the validity and superiority of the proposed method over SOTA methods. 

%Dense light fields have received increasing attention in 3D reconstruction and virtual reality since it can record numerous light rays from a real-world scene. However, current light field representation, such as EPI, sub-aperture images, which are not considered the structure characteristics between the view of the dense light field and consumes a large amount of memory space. What is more, dense light field reconstruction based on these representations requires the disparity or spatial structure information clues. As a result, they have difficulties with occlusions or larger disparity scene. In this paper, we firstly propose a novel Epipolar Focus Spectrum (EFS) representation for dense light fields. Then the characteristics of EFS are analyzed (depth independent, view number dependent). After that, based on these characteristics EFS we proposed a dense light field reconstruction method. The key of the proposed method is  EFS reconstruction by learning the spectrum lines corresponding to the missing view in the frequency domain. This gives our method ability to deal with the reconstruction of the dense light field under different sampling and maintained consistency of views. Experiments on both synthetic and real light fields demonstrate the improvements over state-of-the-arts, especially in large disparity areas.
\end{abstract}

}

\maketitle

\IEEEdisplaynontitleabstractindextext

% \fi
\IEEEpeerreviewmaketitle

\IEEEraisesectionheading{}
\section{Introduction}
\IEEEPARstart{L}IGHT field\cite{adelson1991plenoptic} imaging system records the 3D scene in both the spatial and angular domains \cite{levoy1996light,gortler1996lumigraph}, and has becoming one of the most potential photography techniques for immersive virtual reality \cite{capture4vr2019sig}. However, due to the spatio-angular trade-off\cite{georgiev2006spatio} in the sampling process, it is difficult and expensive to acquire high-resolution light fields, limiting the application and development of light field technologies. Light field reconstruction aims at synthesizing light fields from sparse input views and is an essential tool for generating high-resolution light fields.

% In decades, light field imaging and processing\cite{levoy1996light,ihrke2016principles}, especially dense-view reconstruction, have drawn a lot of attention and gained great progress, 

In decades, dense-view light field reconstruction has drawn a lot of attention and gained great progress, however it still faces many challenging issues. For depth-based methods\cite{eisemann2008floating,goesele2010ambient,chaurasia2011silhouette,chaurasia2013depth,penner2017soft,wanner2014variational,srinivasan2017learning,wang2017light,kalantari2016learning}, the reconstruction results are prone to depth estimation and the view consistency could not be preserved well. For implicit depth-based methods, \textit{i.e.}, the multiplane image (MPI) representation\cite{zhou2018stereo,flynn2019deepview,srinivasan2019pushing,mildenhall2019llff}, the additionally introduced transparency term could not describe intricately occluded areas well (see Fig.\ref{fig:sota_syn}, and Fig.\ref{fig:sota_realdata}).

% the mathematical explanation for the mechanism of the MPI construction has been little explored.

Since the essence of dense-view light field reconstruction is to eliminate the aliasing contents in the Fourier spectrum of the angularly undersampled light field \cite{chai2000plenoptic,zhang2001generalized}, recently several methods have been proposed to focus on recovering the high-frequency spectrum either by modeling the texture consistency in the spatial domain \cite{wang2018end,wu2018light,yeung2018fast,wu2019learning} or inpainting in the transformed domain \cite{shi2014light,vagharshakyan2018light}. However, due to the information asymmetry \cite{yoon2015learning} between the spatial and angular dimensions, the high-frequency spectrum learned or modeled from the light fields with small disparities is inapplicable to the light fields with large disparities, causing artifacts near the occlusion boundaries.

% of reconstructed objects. // The proposed method does not focus on 3D reconstruction, so it is ambiguous to use the term 'reconstructed object'.

%Recently, several methods\cite{shi2014light,wu2017light_epi,vagharshakyan2018light,wu2019learning,wang2018end,yeung2018fast} focus on modelling the texture consistency along different views directly, \textit{i.e.}, super-resolving the epipolar plane image (EPI) in the angular dimension.

In this paper, we define and explore a novel light field representation, called Epipolar Focus Spectrum (EFS) (in Sec.\ref{sec:Epipolar Focus Spectrum Representation for Light Field}). Compared with the classical EPI representation where the slope of the EPI line varies from depth to depth, all contents from the same view will be gathered in a specific line-style zone in the EFS representation,
% the content with the same depth shows the same pattern in the EFS representation, // It is wrong!!!
thus the EFS representation is depth independent. This fundamental characteristic provides the basis for pursuing a unified solution to process full disparity contents simultaneously. 
Apart from this, different from the repeated and overlapped aliasing pattern in the Fourier spectrum of the EPI of an undersampled light field, the EFSs obtained under different angular sampling settings share the same cone-shaped pattern and meet the conjugate symmetry (see Sec.\ref{sec:EFS_characteristics}). Therefore, it is more convenient to accurately reconstruct the dense-view light field using the EFS (frequency domain) representation than the EPI (spatial domain) representation. In Sec.\ref{sec:Dense Light Field Reconstruction Based on EFS}, we first present an end-to-end convolutional neural network (CNN) to eliminate the aliasing contents in the EFS of an undersampled light field, in which a novel conjugate-symmetric loss is adopted for optimization. Then the generated non-aliasing EFS is projected to construct the EPI spectrum. After applying the inverse Fourier transform (IFT) to obtain the dense-view light field, a U-Net with a perceptual loss is finally utilized to optimize the reconstructed results and eliminate the ``trailing image'' \cite{1993A} caused by the integral operation, especially in the marginal view. Experimental results (in Sec.\ref{sec:Evaluations}) verify the effectiveness of the proposed EFS-based dense-view light field reconstruction method.  

The main contributions of the work include,

1) A novel depth-invariant representation for light field is defined, named Epipolar Focus Spectrum (EFS), which guarantees the cross-view consistency for full-depth/disparity light field reconstruction.
% The term EFS has appeared in the previous sections.

2) An important characteristic of EFS is explored, that is a same cone-shaped pattern under different angular samplings due to one-to-one mapping from the view to the EFS line. The pattern is determined by the number of views and the disparity gap between neighboring focal planes.

%\textcolor[rgb]{0.2,0.6,0.4}{2) An important characteristic of EFS is explored. Since there exists a one-to-one mapping from the view to the EFS line, the challenging dense-view reconstruction can be alternatively achieved by reconstructing the linear structures in the frequency domain. Moreover, the aforementioned depth-invariant feature of EFS enables the full-depth/disparity reconstruction process which enforces the cross-view consistency. %EFSs from a light filed with different angular sampling rates share the same cone-shaped pattern and meet the conjugate symmetric (Fig. \ref{fig:differ_samlpy_of_efs}) Therefore, it is more convenient  to  accurately  reconstruct  the  dense-view  light  field using  the  EFSs }

%2) The pattern of EFS is independent of the scene depth, which is  determined by the number of views and the disparity gap ($\Delta\alpha$) between neighboring focal planes. 

%3) An EFS-based learning framework for dense-view light field reconstruction is proposed. Extensive results on both synthetic and real light field datasets verify the superiority of the proposed method, which further validates the advantages provided by the EFS representation.
3) An EFS-based learning framework for dense-view light field reconstruction is proposed. Extensive experiments on both synthetic and real light field datasets verify the superiority of the proposed method.

%Specifically, for the light field reconstruction task which focuses on eliminating the aliasing contents\cite{chai2000plenoptic,vagharshakyan2018light,wu2019pami} in the under-sampled light field, it is 
%
%Additionally, the unique origin and axis symmetries in the EFS . For the light field reconstruction task which focuses on eliminating the aliasing contents\cite{chai2000plenoptic,vagharshakyan2018light,wu2019pami} in the under-sampled light field, it is 

\section{Related Work}
\label{sec:Related Work}

\subsection{Light field representation}
Let $L(u,v,x,y)$ represent the distribution of rays in 3D space, where $(u,v)$ and $(x,y)$ denote the intersections between the ray with angular/camera and spatial/image planes, respectively \cite{levoy1996light, gortler1996lumigraph}. To better model the contents, several representations have been proposed in the literature. In the spatial domain, the sub-aperture image and EPI are the two most common representations. The former emphasizes the spatial information per view. The latter focuses on the disparity among views, where the slope of the EPI lines is associated with the disparity/depth. In the Fourier domain, by exploring the equivalence between the 3D focal stack and a 4D light field, Ng {\it {et al.}} \cite{ng2005fourier} claim the 2D spectrum of a refocused image could be obtained by slicing the corresponding 4D spectrum. Dansereau {\it {et al.}} \cite{dansereau2015linear} propose the hyper-cone and hyper-fan representations, which extend the focal range of each focal slice. Le Pendu {\it {et al.}} \cite{pendu2019fourier} analyse the sparsity of light field spectrum and propose the Fourier disparity layer (FDL) representation considering the spectrum energy concentrates on several slices.  

However, since these representations are highly correlated to the scene depth, the features extracted or learned from the light fields with small disparity range are unsuitable for the light fields with large disparity range, which might leads to wrong inference. In contrast, the proposed EFS is depth-invariant and thus enables an operation or processing covering the whole depth range.

\subsection{Anti-aliasing of refocusing} 
Anti-aliasing requires either abundant angular samples or appropriate filters. This problem has been widely studied in both the spatial and frequency domains. In the spatial domain, Levoy and Hanrahan \cite{levoy1996light} propose a prefilter to reduce the spatial artifacts. Chang {\it {et al.}} \cite{6890175} propose an anti-aliasing method that compensates the effect of undersampling by utilizing depth information. The technique attempts to interpolate more angular samples within each sampling interval using the depth map. Xiao {\it {et al.}} \cite{xiao2017aliasing} further analyze the angular aliasing model in the spatial domain. They first detect the aliasing contents and then use lower-frequency terms of the decomposition to remove the angular aliasing at the refocusing stage. In the frequency domain, Isaksen {\it {et al.}} \cite{isaksen2000dynamically} first propose a frequency-planar light field filtering. Chai {\it {et al.}} \cite{chai2000plenoptic} propose a comprehensive analysis on the trade-off between sampling density and depth resolution. Based on focal stacks and sparse collections of viewpoints, Levin and Durand \cite{levin2010linear} employ the focal manifold in derivations of 2D deconvolution kernels \cite{pendu2019fourier}. After that, Lumsdaine and Georgiev \cite{lumsdaine2008full} discuss the aliasing in terms of the focal manifold and conclude by rendering wide depth-of-field images. By deriving the frequency domain of support of the light field, Dansereau {\it {et al.}} \cite{dansereau2015linear} present a simple, linear single-step filter to achieve volumetric focus effects.

All these methods consider the 3D focal stack as multiple individual slices and remove slice-wise aliasing contents, thus the consistency between neighboring focal slices is not preserved. Different from these methods, the proposed reconstruction method handles aliasing contents of all slices at the same time by treating the 3D focal stack as a whole, so the aliasing could be better removed and meanwhile the consistency is well maintained. 

% All these methods only foone by one, cus on a 2D slice of the 3D focal stack, the aliasing contents are removed slice by slice, so the 
%  Lin et al. \cite{lin2015depth} analyze the symmetry of the light field focal stack in the spatial domain.

\subsection{Dense-view reconstruction}  
As mentioned above, the anti-aliasing operation essentially corresponds to a super-resolution operation in the angular domain \cite{kalantari2016learning,wang2017light}. Existing angular super-resolution methods could be mainly divided into two categories. 

The first category is based on depth estimation \cite{goesele2010ambient,chaurasia2013depth,penner2017soft,srinivasan2017learning}. Wanner {\it {et al.}} \cite{wanner2014variational} reconstruct novel viewpoints by combining input viewpoints and estimated depth information. In order to synthesize novel views from a sparse set of input views, Kalantari {\it {et al.}} \cite{kalantari2016learning} propose two convolutional neural networks to estimate the depth and color of each viewpoint sequentially. %In order to generate a viewpoint map of arbitrary viewpoint, Kalantari et al. \cite{kalantari2016learning} propose a parallax estimation depth convolutional neural network (CNN) and color estimation depth CNN to estimate depth value and color of each viewpoint, respectively.
Srinivasan {\it {et al.}} \cite{srinivasan2017learning} present a machine learning algorithm that takes a 2D RGB image as input and synthesizes a 4D RGBD light field. Specifically, their pipeline consists of a CNN that estimates scene geometry, and a second CNN that predicts occluded rays and non-Lambertian effects. Subsequently, Srinivasan {\it {et al.}} \cite{srinivasan2019pushing} propose to utilize the MPI representation to synthesize the viewpoint from a narrow baseline stereo pair. 

The second category focuses on modeling the consistency of EPI texture \cite{wu2018light,yoon2015learning,guo2018dense,zhu2019tvcg} or the sparsity of Fourier spectrum \cite{levin2010linear,shi2014light,vagharshakyan2018light}. Wu {\it {et al.}} \cite{wu2018light} convert the angular domain reconstruction of a 4D light field into a one-dimensional super-resolution of the 2D EPI. Zhu {\it {et al.}} \cite{zhu2019tvcg} further improve the super-resolution performance on EPI in large disparity areas by introducing a long-short term memory module. Considering the special 2D mesh sampling structure of the 4D light field, Levin {\it {et al.}} \cite{levin2010linear} utilize the 3D focus stack to complement the spectrum of the 4D light field and achieve the dense reconstruction for sparse light fields. Shi {\it {et al.}} \cite{shi2014light} exploit the sparsity of the 4D light field in the continuous Fourier domain and perform the dense light field reconstruction by adopting the sparse Fourier transform. Vagharshakyan {\it {et al.}} \cite{vagharshakyan2018light} utilize a sparse representation of the underlying EPIs in shearlet domain and employ an iterative regularized reconstruction.

%{\textcolor{red}{
%Considering the relationship between the aliasing phenomenon of light field refocusing and view sampling rate, unlike previous methods, we analyze the focus stack slice in the frequency domain and observe frequency characteristics under different sampling rates. Considering the depth estimation may be noisy and inaccurate, the view reconstruction results obtained by the above mentioned methods tend to be distorted. What is more, 
%angular super-resolution based on the frequency domain operations and EPI can not deal with light field reconstruction with large parallax. Our method does not require depth estimation and resolve the large disparity problem by mapping it into a limited bandwidth frequency space of the focus stack slice.
%}}
Nevertheless, these methods either rely on accurate depth calculations (especially near occlusion boundaries) or are inappropriate for large disparity scenes since the required texture lines or sparse spectrum features are not available. Differently, based on the depth-independent EFS representation, the proposed light field reconstruction method avoids the challenging depth estimation and optimize the contents at various depths with the same strategy.

%-------------------------------------------------------------------------
\section{The EFS Representation}
\label{sec:Epipolar Focus Spectrum Representation for Light Field}
In this section, we first define the EFS representation for a light field, then introduce its depth-independent characteristics. %To better describe the EFS, 2D EPI representation $E(u,x)$ is used in this section.

\subsection{Notations}
\label{subsection:Notations}
For better understanding the definition of EFS, we first list the notations used in this work, as listed in Tab.\ref{tab:notations}.

Given a 4D light field $L(u,v,x,y)$, $E(u,x)$ denotes its EPI when fixed $v=v^*$ and $y=y^*$ and $F(f,x)$ denotes the 2D focal stack integrated by $E_{d}(u,x)|_{d=f}$, where $(u,v)$ and $(x,y)$ refer to the angular and spatial dimensions respectively. $FT_{*}(\cdot)$ and $FT_{2D}(\cdot)$ denote the 1D and 2D Fourier transform respectively (* means specific variable). $u_{ref}$ is the reference view. 

In brief, we have the following two equations, 
\begin{enumerate}
    \item $\mathcal{E}(\omega_{u},\omega_{x}) \triangleq FT_{2D}(E(u,x))$. 
    \item $EFS(\omega_{f},\omega_{x}) \triangleq FT_{f}(\mathcal{F}(f,\omega_{x}))$.
\end{enumerate}

%$\mathcal{E}(\omega_{u},\omega_{x})$ is the Fourier spectrum of $E(u,x)$, \textit{i.e.}, $%\mathcal{E}(\omega_{u},\omega_{x})=|FT_{2D}(E(u,x))|$. 
%$\mathcal{F}(\omega_{f},\omega_{x})$ characterizes the 2D spectrum of $F(f,x)$.
%$\mathcal{F}(f,\omega_{x})$ characterizes the 1D Fourier spectrum of the $F(f,x)$ along the $x$-axis. $FT_{f}(\cdot)$ applies 1D Fourier transform to $F(f,x)$ along the $f$-axis. 
%2) $\mathcal{F}(f,\omega_{x}) \triangleq FT_{x}(F(f,x))$.

%$FT_{2D}(\cdot)$ applies 2D Fourier transform to $F(f,x)$. 

% $d$ characterizes the disparity during the shearing process\cite{ng2006digital}.

\begin{table}[t]
	\small
	\begin{center}
		\caption{Related notations of the EFS representation.}
		\vspace{-0.2cm}
		\label{tab:notations}      % Give a unique label
		\begin{tabular}{ll}
			\hline\hline
			Term & Definition\\ \hline
			$L(u,v,x,y)$ & A 4D light field \\
            $u,v$ & Angular coordinates \\
            $x,y$ & Spatial coordinates \\
			$E(u,x)$ & 2D EPI \\
			$E_d(u,x)$ & Sheared EPI at the specific disparity $d$ \\
    		$d_{range}$      & Disparity range for the shearing process \\
    		$N_f$      & The number of refocus layers \\
			$FT_{*}(\cdot)$ & 1D Fourier transform on the variable * \\
			$FT_{2D}(\cdot)$ & 2D Fourier transform  \\
			$F(f,x)$ & Focal stack integrated by $E_{d}(u,x)|_{d=f}$ \\
			$\mathcal{E}(\omega_{u},\omega_{x})$ & Fourier spectrum of $E(u,x)$ \\
			$\mathcal{F}(f,\omega_{x})$ & Slicing and rearranging of $\mathcal{E}(\omega_{u},\omega_{x})$\\%& 1D transform of $F(f,x)$ on $x$, $FT_{x}(F(f,x))$ \\
			%$\hat{F}(\omega_{d},\omega_{x})$ & 2D spectrum of $F(d,x)$ \\
%			\multirow{2}{*}{$FT_{f}(\cdot)$} & 1D Fourier transform of $F(f,x)$ along the \\ &$f$-axis \\
%			\multirow{2}{*}{$EFS(\omega_{f},\omega_{x})$}
			
%             & Epipolar focus spectrum or the\\ & 2D Fourier spectrum of the $F(f,x)$\\
            $EFS(\omega_{f},\omega_{x})$ & EFS or 2D Fourier spectrum of $F(f,x)$\\ 
			$u_{ref}$ &  Reference view (center view in this work) \\
			\hline\hline
		\end{tabular}
	\end{center}
	
\end{table}
\vspace{-0.5cm}

\subsection{Definition and Construction of EFS}
\label{sec:efs_definition}
In this section, we first give the definition of EFS and then present two methods for constructing EFS.

\noindent \textbf{Definition.} Given a 2D EPI $E(u,x)$, the EFS representation is defined as the 1D Fourier transform along the $f$-axis of the rearranged EPI Fourier spectrum $\mathcal{E}(\omega_{u},\omega_{x})$ slices. 
In detail, a slice operation is firstly applied on $\mathcal{E}(\omega_{u},\omega_{x})$ to construct $\mathcal{F}(f,\omega_{x})$, %Eq.\ref{eqn:fs_x_fourier} is the slice  operation of EPI Fourier spectrum on different disparity $f$,
\begin{equation} \small
	\mathcal{F}(f,\omega_{x})=\mathcal{E}(-f\omega_{x},\omega_{x}),
	\label{eqn:fs_x_fourier}
\end{equation}
where the disparity takes value in the interval $f\! \in\! [ d_{min}, d_{max}] $.
%Then, we rearranged these slice followed by the range of disparity. Finally, 
Then the EFS is obtained by the 1D Fourier transform of $\mathcal{F}(f,\omega_{x})$ along the $f$-axis,
\begin{equation} \small
	EFS(\omega_{f},\omega_{x}) = FT_{f}(\mathcal{F}(f,\omega_{x})).
	\label{eqn:fs_xd_fourier}
\end{equation}

According to Eq.\ref{eqn:fs_x_fourier}, $\mathcal{F}(f,\omega_{x})$ contains all spectrum contents of the EPI when $d_{min}\rightarrow -\infty$ and  $d_{max}\rightarrow +\infty$. In other words, the EFS could represent the EPI losslessly in this case.
%representationEFS is equivalent to EPI spectrum in this case. In other words, the  of light field is lossless.

\begin{figure}[tb]
\begin{center}
\centering
\includegraphics[width=\linewidth]{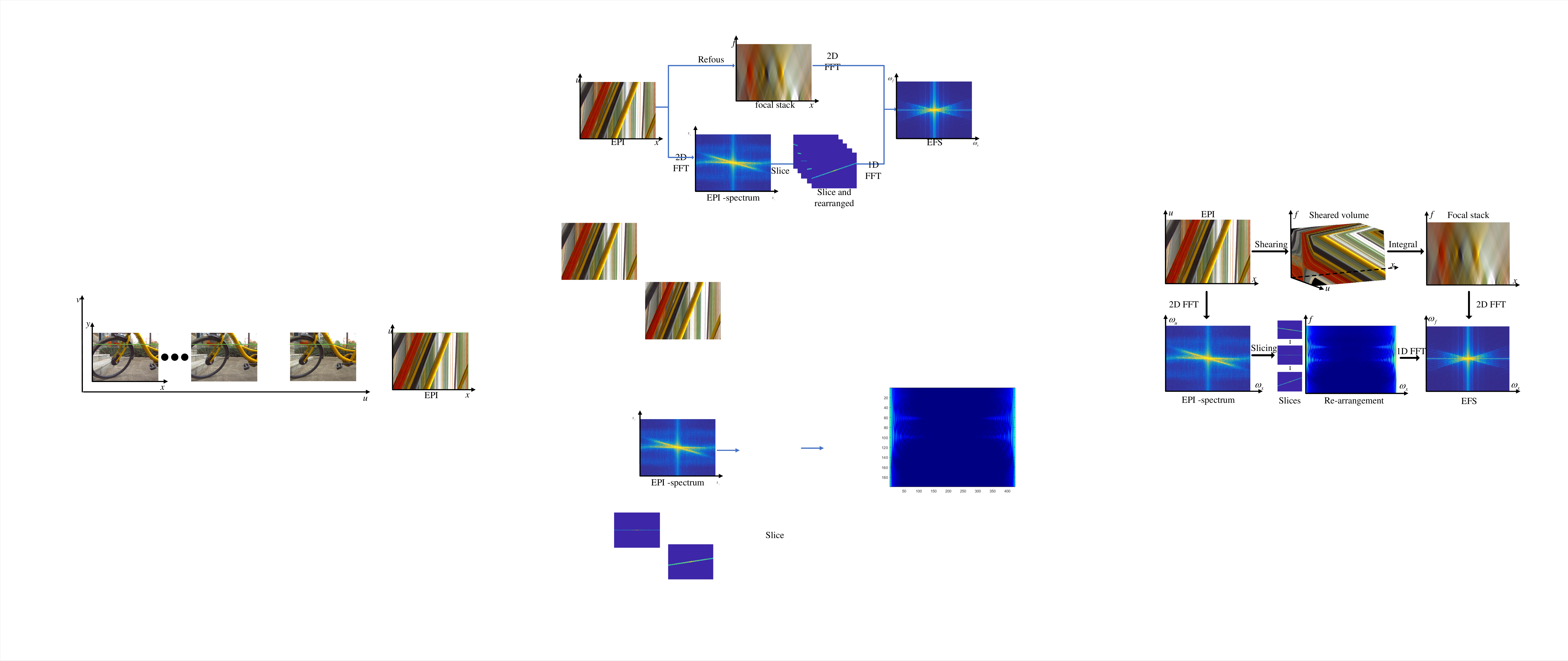}
\end{center}
\vspace{-0.3cm}
\caption{Two different ways to obtain the EFS, either via the focal stack (top flow) or via the EPI spectrum (bottom flow)}.
\vspace{-0.3cm}
\label{fig:define_EFS}
\end{figure}
\vspace{0.5cm}

\noindent \textbf{Construction.} From the above definition, the EFS could be constructed based on the Fourier slice of $\mathcal{E}(\omega_{u},\omega_{x})$, as shown in the bottom branch in Fig.\ref{fig:define_EFS}.
% Eq.\ref{eqn:fs_x_fourier} and Eq.\ref{eqn:fs_xd_fourier}. The  bottom flow of Fig.\ref{fig:define_EFS} is also shows the constructed of EFS based on Fourier slice. 

%The second method is to obtain EFS by focal stack in spatial domain. According to the Fourier slice theorem\cite{ng2005fourier} , the EFS equals to the 2D Fourier spectrum of the focal stack, so the EFS could  be constructed in the spatial domain(the top flow in Fig.\ref{fig:define_EFS}). As shown in Eq.\ref{eqn:epi_shearing}, we first shearing the EPI at specific disparty $d$,

%According to the Fourier slice photography theorem\cite{ng2005fourier}, the above slice opertion applied to the $\mathcal{E}(\omega_{u},\omega_{x})$ is equivalent to the shearing operation in the spatial domain EPI, thus the EFS could also be constructed from the spatial domain. Firstly, a shearing opertion is applied,
%\begin{equation}
%E_f(u,x) = E(u,f(u-u_{ref}))
%\end{equation}
%
%\begin{equation} \small
%\begin{aligned}
%&E_f(u,x) = E(u,f(u-u_{ref}))  \\
%&F(f,x) = \int E_f(u,x)\mathbf{d}u=\int E(u,x+f(u-u_{ref})\mathbf{d}u\\
%&EFS(\omega_{f},\omega_{x}) = FT(F(f,x)).
%\end{aligned}
%\label{eqn:EFS__from_space}
%\end{equation}

Alternatively, %following the above Fourier domain based construction method, 
the EFS could also be obtained in the spatial domain according to the Fourier slice photography theorem \cite{ng2005fourier}. Firstly,  $E_d(u,x)$ denotes the sheared EPI at a specific disparity $d$, 
\begin{equation}\small
  E_d(u,x)=E(u,x+d(u-u_{ref})).
\label{eqn:epi_shearing}
\end{equation}
%where $u_{ref}$ refers to the reference view.  

After that, once the sheared EPI with an arbitrary disparity $f \in \left[ d_{min}, d_{max}\right] $ is integrated over all the views, the focal stack $F(f,x)$ is formed,
\begin{equation}\small
 F(f,x) = \int E_f(u,x){d}u=\int E(u,x+f(u-u_{ref})){d}u.
 \label{eqn:focaltack}
\end{equation}

Then by performing the 2D Fourier transform on the focal stack, we can obtain the corresponding EFS representation,
\begin{equation}\small
EFS(\omega_{f},\omega_{x}) \triangleq FT_{2D}(F(f,x)).
\label{eqn:EFS_from_space}
\end{equation}

Noting that, the lossless EFS could only be constructed in two cases, \textit{i.e.}, $\{d_{min},d_{max}\} \rightarrow \{-\infty,+\infty\}$ or the infinite aperture size \cite{levin2010linear}. However, since these conditions are practically impossible to achieve, the EFS is actually a lossy representation in practice. To minimize the effects of the missing spectrum in $\mathcal{E}(-f\omega_{u},\omega_{x})$, it is suggested to set $d_{min}$ and $d_{max}$ as the minimum and maximum disparities of the scene respectively. A detailed analysis on this issue will be given in Sec.\ref{sec:Ablation Studies} .

%Following the above Fourier domain based construction method, the EFS could also be obtained from the spatial domain of EPI according to the Fourier slice photography theorem\cite{ng2005fourier}. 
%\begin{equation}\small
%  E_d(u,x)=E(u,d(u-u_{ref}))
%\label{eqn:epi_shearing}
%\end{equation}
%Once the sheared EPI   with arbitrary disparity $f \in \left[ d_{min}, d_{max}\right] $ is integrated all views, the focal stack $F(f,x)$ is formed,
%\begin{equation}\small
% F(f,x) = \int E_f(u,x)du=\int E(u,x+f(u-u_{ref})du.
% \label{eqn:focaltack}
%\end{equation}
%Then perform Fourier transform on focal stack to get EFS,
%\begin{equation}\small
%EFS(\omega_{f},\omega_{x}) = FT(F(f,x)).
%\label{eqn:EFS_from_space}
%\end{equation}

%According to \ref{eqn:fs_x_fourier}, when $d_{min}\rightarrow +\infty, d_{max}\rightarrow -\infty$,  given a EFS, we can construct the light field spectrum. In other words, the EFS representation for light filed is lossless.
%
%According to\cite{levin2010linear}, for infinite aperture, if the shearing operation of EPI or  slice  operation of  $\mathcal{E}(\omega_{u},\omega_{x})$ can take all the disparity of the scene, combine the above derivation, given a EFS, we can construct the light field spectrum. In other words, the EFS representation for light filed is lossless.

% However, there is no infinite aperture  camera at the moment, so  EFS representation for light filed has certain loss. When aperture is fix,为了验证，损失上限，我们统计了10000张由包含场景所有深度范围的EFS重建而来的EPI平铺能量损失，如图//所示，其损失基本集中在5%以内。
\begin{figure*}[!tb]
\centering
\includegraphics[width=0.9\linewidth]{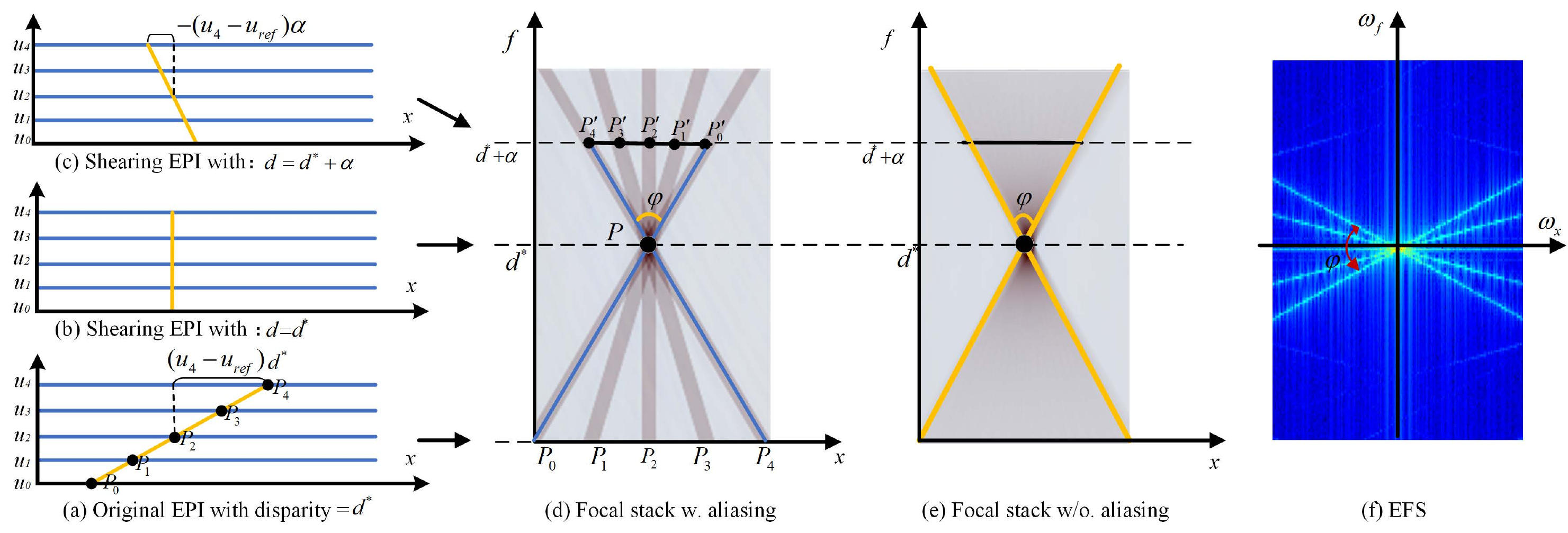}	
\vspace{-0.3cm}
\caption[example]
{
  Analysis for the EPI in the $u$-$x$ space, the focal stack in the $f$-$x$ space and the EFS in the $\omega_f$-$\omega_x$ space. (a)-(c) Original and sheared EPIs with different parameters, where $u_{ref}=u_2$. (d) Aliased focal stack. (e) Continuous focal stack. (f) Fourier spectrum of (d). For better visualization, only one single pixel $P$ is considered here. \label{anysis_EFS}
}
\vspace{-0.3cm}
\end{figure*}
\vspace{-0.5cm}

\subsection{Characteristics of EFS}	
\label{sec:EFS_characteristics}
According to the second construction method, the EFS is equal to the Fourier spectrum of the focal stack. In this section, we first analyze the characteristics of the focal stack under different values of $f$, then discuss the attributes of EFS in the Fourier domain.

\subsubsection{Defocus and Aliasing in the Focal Stack}
Fig.\ref{anysis_EFS} (a) shows an EPI line for a pixel with disparity $d^*$. The EPI is sheared with $d^*$ (in Fig.\ref{anysis_EFS} (b)) according to Eq.\ref{eqn:epi_shearing}. Since the EPI line is now perpendicular to the $x$-axis, there is no aliasing or defocus blur at $P$ in the refocused image, as shown in Fig.\ref{anysis_EFS} (d) and (e). Then the original EPI is sheared with $d^*+\alpha$ (see Fig.\ref{anysis_EFS} (c)). It is noticed that the EPI line is not perpendicular to $x$-axis anymore and the aliasing or defocus blur appears again in the focal stack, as illustrated in the layer $d^*+\alpha$ in Fig.\ref{anysis_EFS} (d) and (e) respectively. The radius of the aliasing or defocus blur increases with the increasing of $\alpha$ and a triangle is formed in the focal stack slice. Noting that, since the baseline between neighboring views is scaled by $\frac{1}{M+1}$ when inserting $M$ views, the defocus diameter in Fig.\ref{anysis_EFS}(e) is equal to the aliasing distance in Fig.\ref{anysis_EFS}(d). The apex angle $\varphi$ in Fig.\ref{anysis_EFS}(e) does NOT change (the same cone-shaped pattern still exists, as shown in Fig.\ref{anysis_EFS}(e)).

\begin{figure}[!tb]
\begin{center}
\centering
\includegraphics[width=0.95\linewidth]{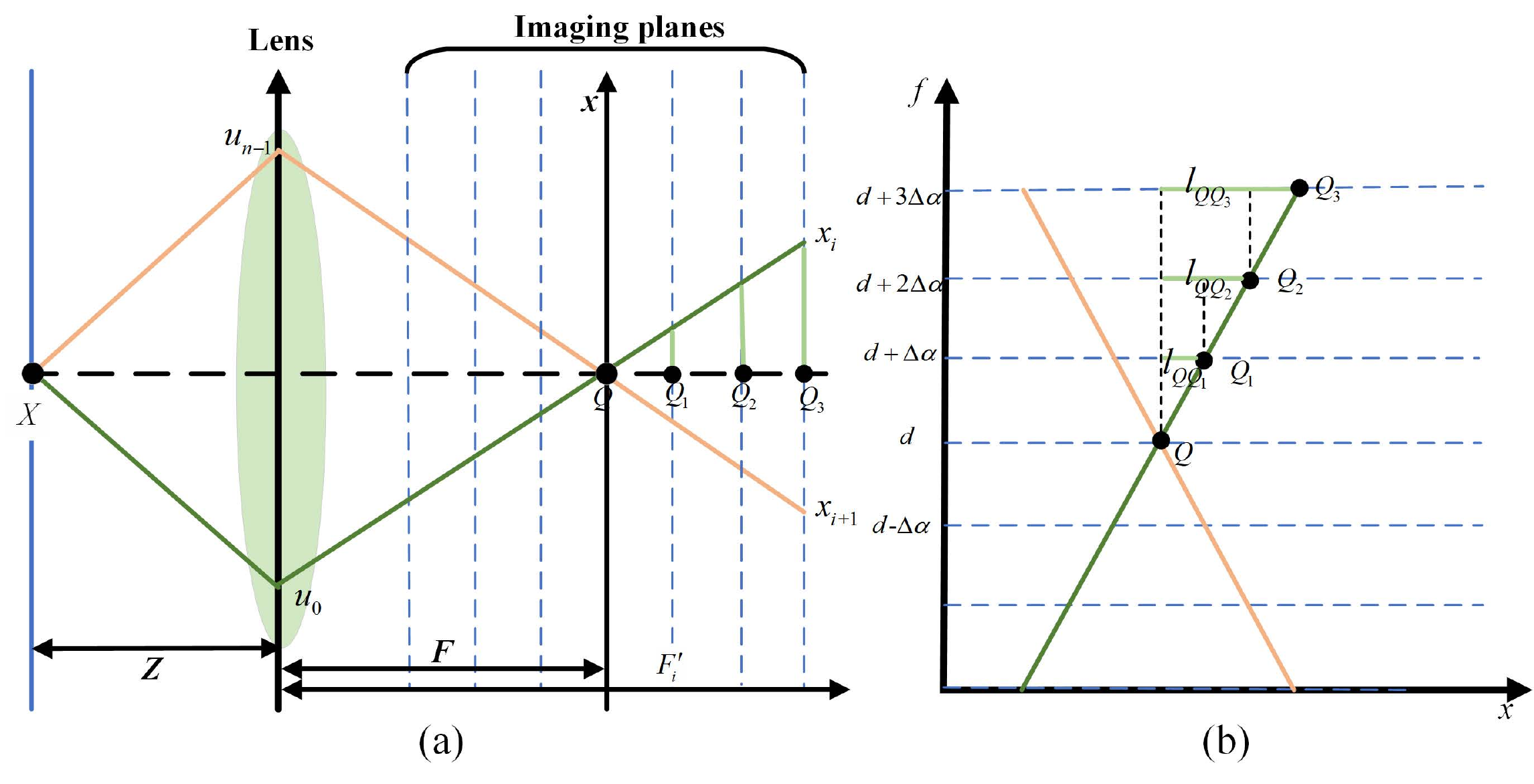}
\end{center}
\caption{EFS sampling analysis. (a) Optical path of the point $X$ in the focal stack. (b) Schematic diagram of the corresponding focal stack.}
\label{fig:Sampling_analysis_of_EFS}
\vspace{-0.5cm}
\end{figure}

Without loss of generality, only the horizontal angular sampling is discussed here. Suppose that the light field has $N_u$ views, the apex angle $\varphi$ of the triangle has two forms,
\begin{equation}
 \small
\label{eqn:efs_angle_continuous}
\varphi \!=\!
\begin{cases} 
2\arctan\! \left(\frac{1}{2}(N_u-1)\right) & \text{Continuous focal stack}\\
2\arctan\! \left(\frac{1}{2}\Delta\alpha(N_u-1)\right) & \text{Discrete focal stack}
\end{cases}
,
\end{equation}
where $\Delta\alpha$ is the disparity gap between neighboring focal planes when constructing the focal stack as shown in Fig.\ref{fig:Sampling_analysis_of_EFS}.

The diameter $l$ of the defocus blur (or aliasing), \textit{i.e.}, the interval between $P_0$ and $P_4$ in Fig.\ref{anysis_EFS} (d), can be calculated by
\begin{equation}
\label{eqn:defocus_diameter}
\small
l_{P_0,P_4} = |\alpha(N_u-1)|.
\end{equation}
Note that, the aliasing only occurs when $|\alpha(N_u-1)|>N_u-1$ holds. Furthermore, for textured areas, the aliasing always exists in the refocused image as long as the sheared parameter $|\alpha|$ is large enough.

Revisiting Eq.\ref{eqn:epi_shearing},
it is found that the aliased pixels $P_0$, $...$, $P_4$ come from the views $u_0$, $...$, $u_4$ respectively. The slope of each aliasing line $PP_i$ in Fig.\ref{anysis_EFS}(d) can be computed by
\begin{equation} \small
\label{eqn:slope_aliasing_line}
Slope(PP_i)=
\begin{cases}
\frac{1}{u_i-u_{ref}} & \text{Continuous focal stack}\\
\frac{1}{\Delta\alpha(u_i-u_{ref})} & \text{Discrete focal stack}
\end{cases}
.
\end{equation}

According to Eqns.\ref{eqn:efs_angle_continuous} and \ref{eqn:slope_aliasing_line}, two conclusions are drawn, % in the focal stack slice, %{\textit{ the shape of defocus blur or aliasing line is determined by the refocus parameter $\Delta\alpha$ and the view index, and is \textbf {independent} of the scene depth. \textbf{All aliasing lines from the same view have the same slope in the focal stack.}}}

\begin{enumerate}[a)]
  \item The shape of the defocus blur or aliasing line is determined by both the disparity gap $\Delta\alpha$ and the view index, and it is independent of the scene depth.
  \item All aliasing lines from the same view have the same slope.
\end{enumerate}

\subsubsection{Views sampling in the EFS}
\label{subsection:Views sampling in EFS}
According to the second property, given a focal stack formed from a $N_u$-view light field, there are $N_u$ frequency lines in the EFS according to the property of the Fourier transform \cite{bigun1987optimal}\footnote{Fourier transform tells the energy of all lines with the same slope in the spatial domain concentrates on a perpendicular line passing through the origin.}. Each line corresponds to a specified view, and more views lead to more lines. Consequently, the EFS has the following properties,
\begin{enumerate}[a)]
  \item The shape of the EFS pattern is determined by both the refocus parameter $\Delta\alpha$ and the number of views, and it is independent of the scene depth.
  \item According to the property of the Fourier transform \cite{gonzales2002digital}, the EFS is conjugate symmetric. 
\end{enumerate}

\begin{figure}[t]
\begin{center}
\centering
\subfloat[]{
	\includegraphics[width=1in]{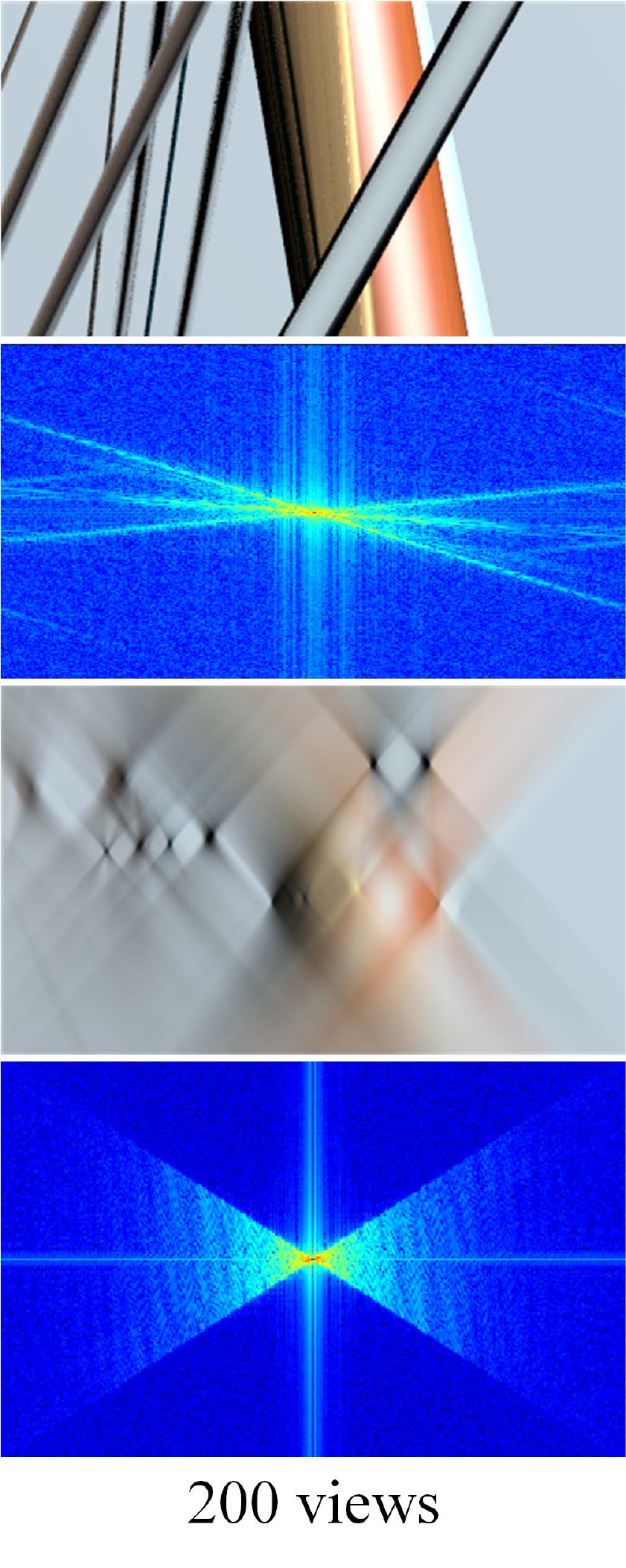} %53, 70
}
\subfloat[]{
	\includegraphics[width=1in]{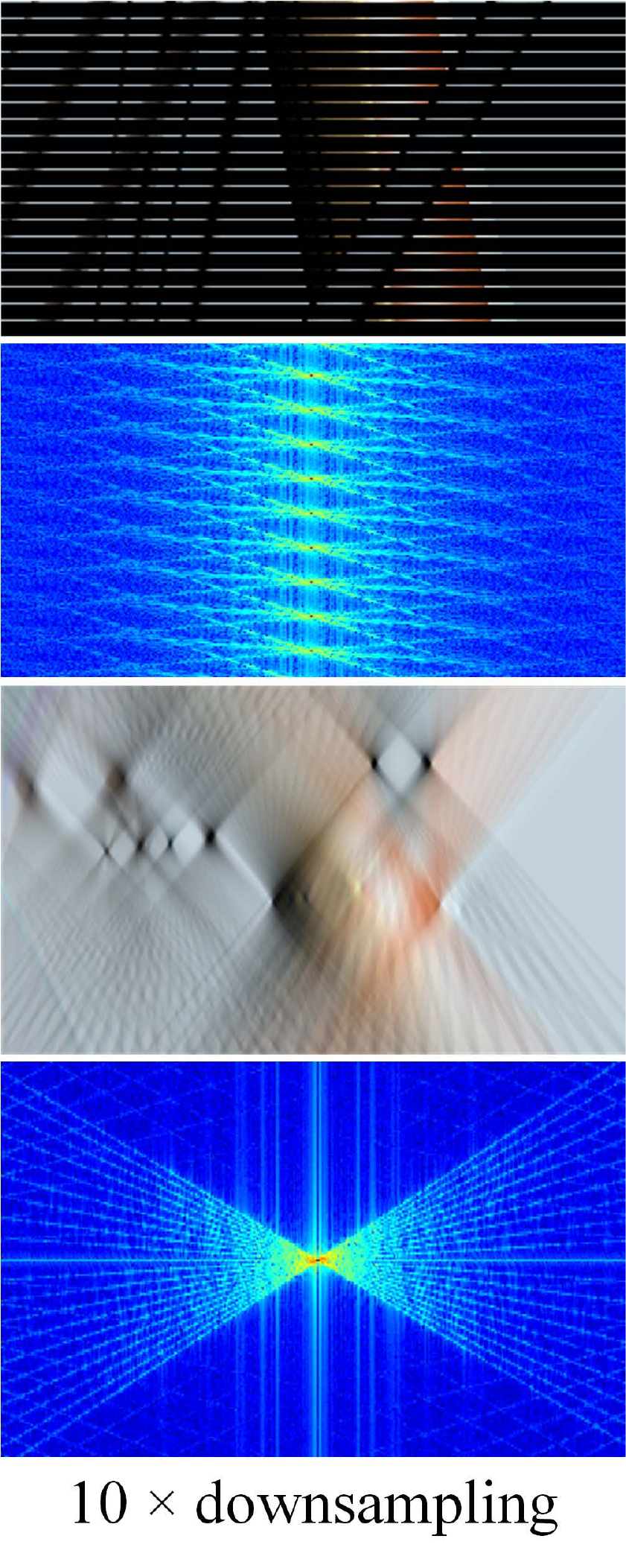} %56.3, 72.3
}
\subfloat[]{
	\includegraphics[width=1in]{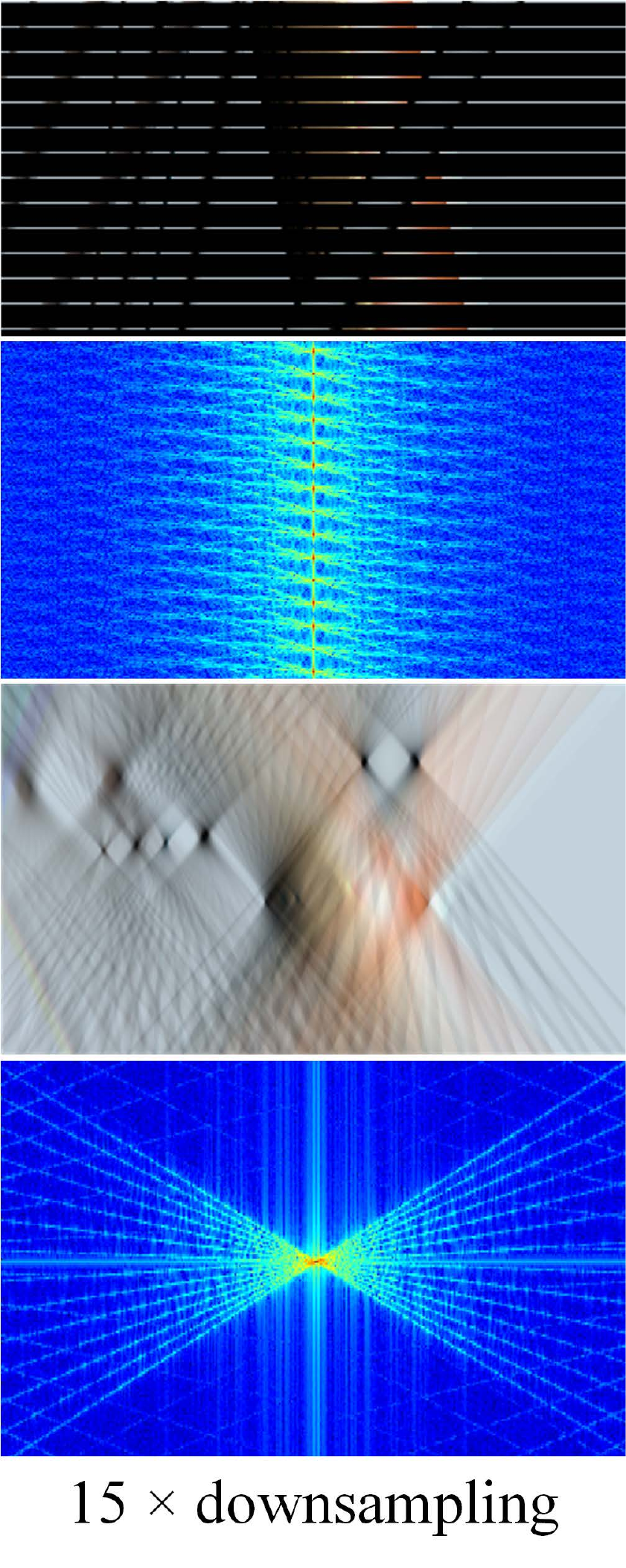} %56.3, 72.3
}
\end{center}
\vspace{-0.3cm}
\caption{Illustrations of the EPI and EFS under different downsampling rates. From top to bottom: EPI, EPI spectrum, focal stack and EFS.}
\label{fig:differ_samlpy_of_efs}
\vspace{-0.4cm}
\end{figure}
%\vspace{-0.5cm}

Fig.\ref{fig:differ_samlpy_of_efs} provides several examples of the EPI, EPI spectrum, focal stack and EFS under different sampling rates. Taking a closer look at the EPI spectrum (the 2nd row) and the EFS (the last row) in Fig.\ref{fig:differ_samlpy_of_efs}, it is worth noting that with a reduction of the view count, more repeating patterns appear in the EPI spectrum, while the structure of the EFS distribution almost remains unchanged. Additionally, there is a one-to-one correspondence between the line in the EFS and the view index when $\Delta\alpha$ is fixed. As shown in Fig.\ref{fig:differ_samlpy_of_efs}(c), 14 views in the EPI correspond to 14 lines in the EFS.

\begin{figure}[!tb]
\begin{center}
\centering
\includegraphics[width=0.85\linewidth]{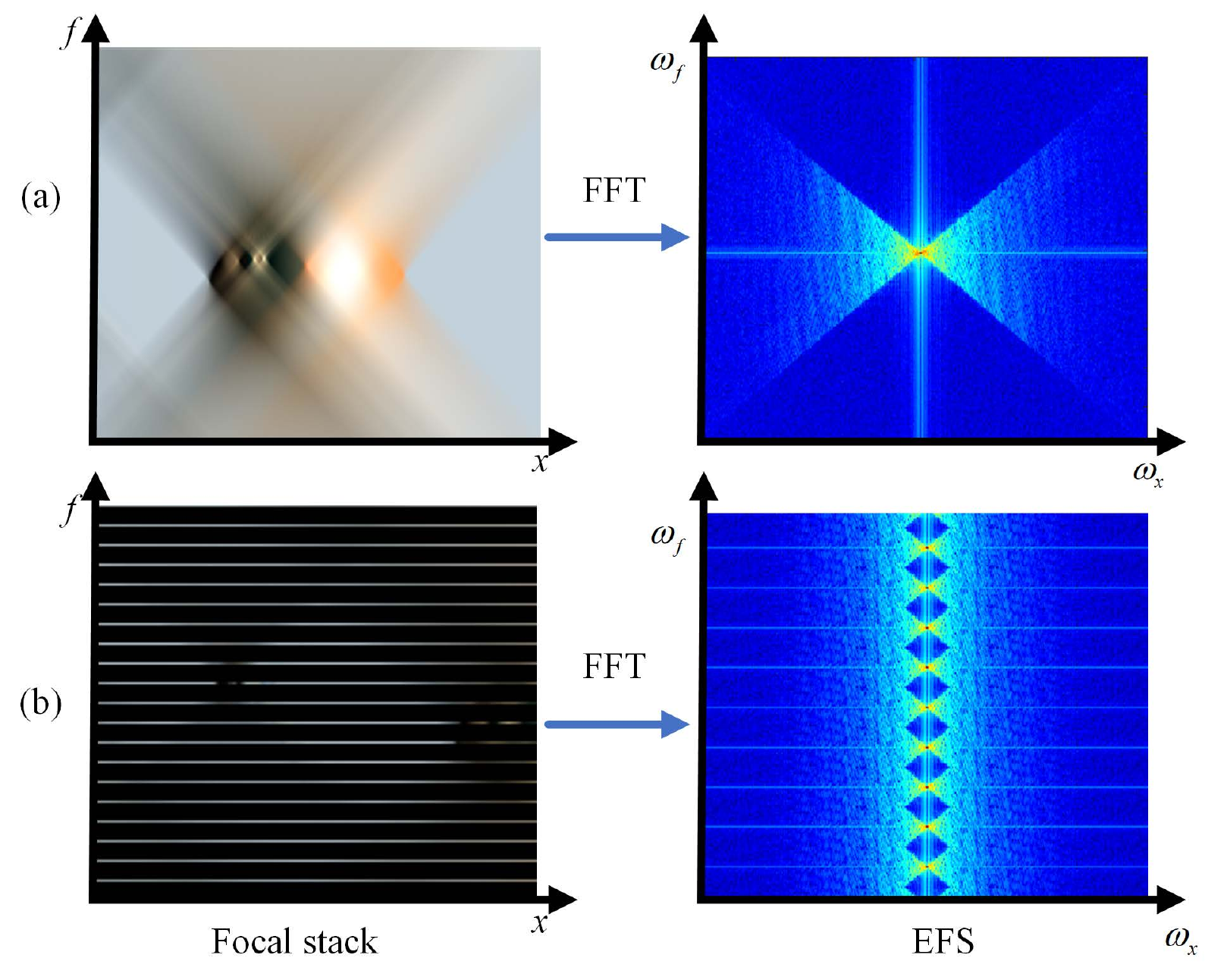}
\end{center}
\vspace{-0.3cm}
\caption{Illustration of EFSs under different focal layer counts. (a) Sufficient focal layers result in a non-aliased EFS. (b) Insufficient focal layers result in an aliased EFS. }
\label{fig:undersamplingEFS}
\vspace{-0.3cm}
\end{figure}

\begin{figure}[!t]
\begin{center}
\centering
\includegraphics[width=0.68\linewidth]{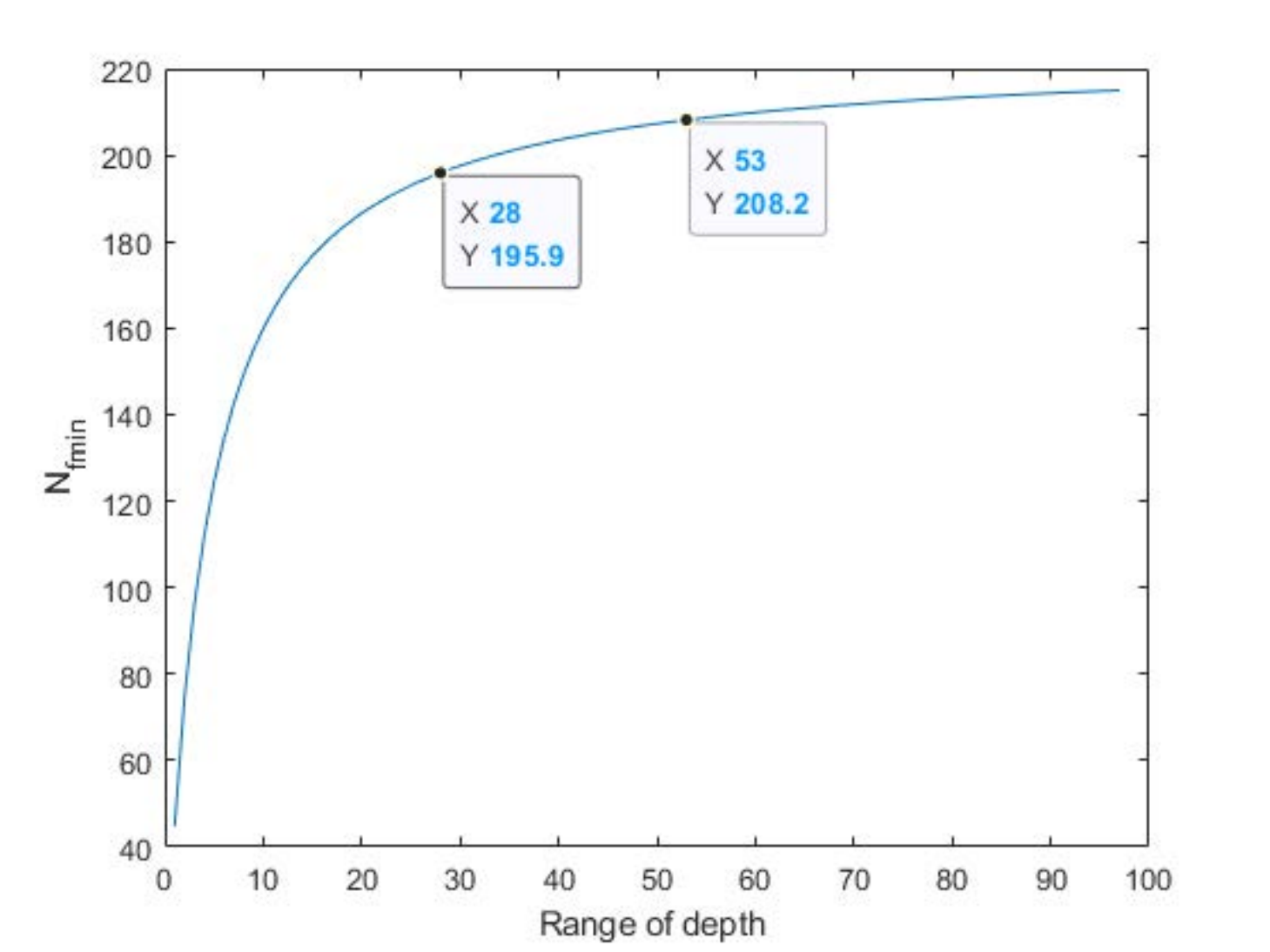}
\end{center}
\vspace{-0.3cm}
\caption{The varying trend of $N_{fmin} $ regarding $Z_{min}$ and depth range. }
\label{fig:N_cmin}
\vspace{-0.6cm}
\end{figure}

\begin{figure*}[!h]
\begin{center}
\centering
\includegraphics[width=\linewidth]{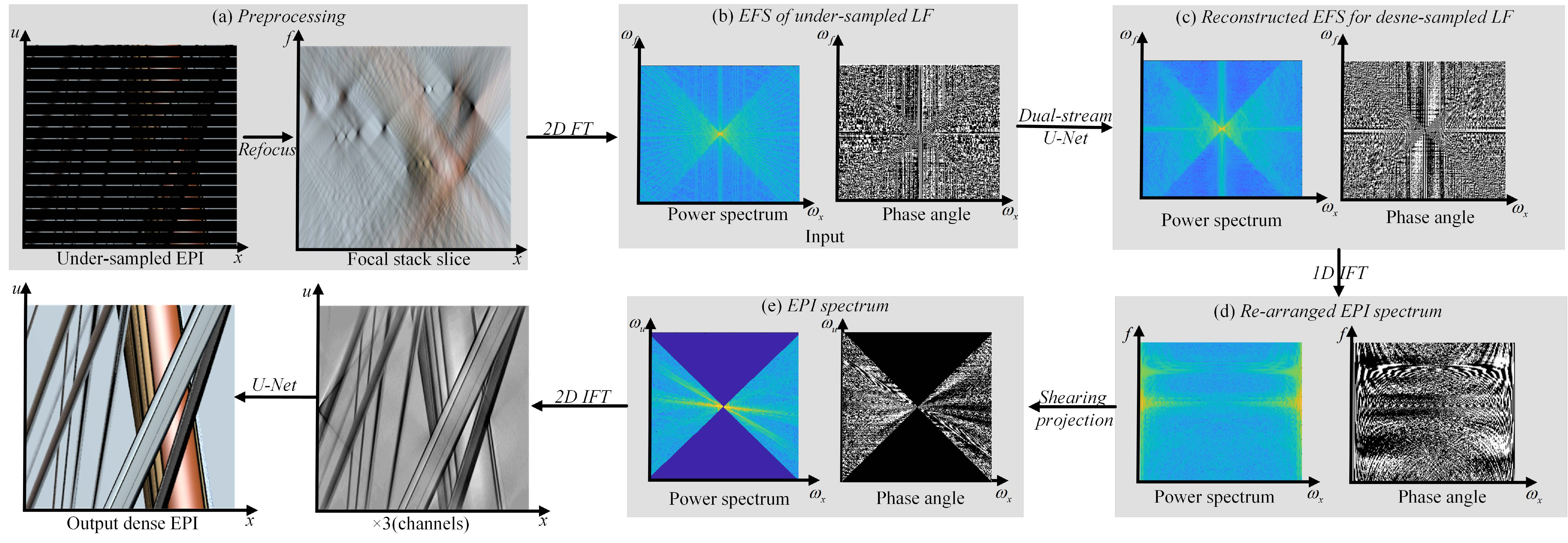}
\end{center}
\vspace{-0.3cm}
\caption{The pipeline of the proposed EFS-based dense-view light field reconstruction. The whole pipeline consists of preprocessing, EFS reconstruction, shearing projection and final optimization of the reconstructed EPI.}
\label{fig:pipline}
\end{figure*}

\vspace{-0.2cm}
\subsection{Sampling analysis of EFS}
\label{sec:Sampling analysis of EFS}

There are mainly two reasons for aliasing in the EFS, one is sparse angular sampling (Fig.\ref{fig:differ_samlpy_of_efs}(b)(c)), the other is insufficient refocus layers (Fig.\ref{fig:undersamplingEFS}). Since the EFS structure basically remains in the first case, here we focus on the second case caused by insufficient refocus layers, and analyze the lower bound of focal stack layers required for non-aliasing EFS recovery. 
%%%%%%%%%%%%%%%%%%%%%%
%需要加上一个说明：当聚焦深度范围等于场景中所有物体深度范围时。

As shown in Fig.\ref{fig:undersamplingEFS}, when the refocus range of the focal stack is equal to the depth range for all the objects in the scene, aliasing appears when the disparity gap $\Delta \alpha$ between neighboring focal layers increases. To eliminate aliasing, all lines formed from different views in the focal stack ought to be continuous instead of being discrete, \textit{i.e.}, $\Delta \alpha(u_i-u_{ref})\leq 1, ~\forall i\in [1,N_u]$. All the inequalities for the middle views hold if the inequalities for two marginal views hold. In other words, the disparity gap $\Delta \alpha$ should meet the following inequality,
\begin{equation} \small  \label{eqn:range_of_layer_alpha}
\Delta \alpha\frac{N_u-1}{2} \leq 1.
\end{equation} 

The refocus range $[d_{min}, d_{max}]$ is set according to the scene depth range $[Z_{min}, Z_{max}]$,
 \begin{equation} \small  \label{eqn:corresponding_disparity}
d_{max}=\frac{kB}{Z_{min}}, \: d_{min}=\frac{kB}{Z_{max}},
\end{equation}
where $k$ is the focal length and $B$ is the baseline. Thus, the number of refocus layers $N_f$ in the focal stack is calculated by
\begin{equation}\small  \label{eqn:focal_stack_layers} 
N_f
=\frac{d_{max}-d_{min}}{\Delta \alpha}
=\frac{kB(Z_{max}-Z_{min})}{\triangle\alpha(Z_{min}Z_{max})}.
\end{equation}

Combining Eqns.\ref{eqn:range_of_layer_alpha} and \ref{eqn:focal_stack_layers}, we have %$N_f$ meets,
\begin{equation} \small \label{eqn:range_of_layer}
N_f \geq \frac{kB(Z_{max}-Z_{min})(N_u-1)}{2Z_{max}Z_{min}}.
\end{equation}

When the depth is discontinuous, it is essential to take the scene distribution into consideration. The minimum number of focal layers is estimated as
\begin{equation} \small \label{eqn:final_range_of_layer}
N_{fmin}= S(Z,O,T)\frac{kB(Z_{max}-Z_{min})(N_u-1)}{2Z_{max}Z_{min}},
\end{equation}
where $S(Z,O,T)$ denotes the scene distribution function, determined by the depth $Z$, occlusion $O$ and texture $T$.

In summary, the lower bound of focal stack layers is determined by the relative depth variation ($Z_{min},Z_{max} $), scene distribution $S(Z,O,T)$ and the baseline $kB$ of the synthetic aperture optical system. Fig.\ref{fig:N_cmin} illustrates an example of choosing $N_{fmin}$, where $N_{u}=200$, $S(Z,O,T)=1$ (a continuous depth distribution), $kB=9$, $Z_{min}\in[2,10]$. The depth takes value in the interval $[1,100]$. Particularly, the black line curve shows the varying trend of $N_{fmin} $ with $Z_{min}=4$. Refer to Sec.\ref{sec:Ablation Studies} for more experimental analysis.

%On our synthetic dataset,  we set $kB=9$ and the number of dense light field viewpoint is $200$, supposing the depth distribution is continuous ( $S(Z,O,T)=1$ ), the $Z_{min}$ is 4, and the  ranges of  $Z_{max}$ is $5$ to $100$, the variation trend of $N_{fmin} $ will shown in Fig.\ref{fig:N_cmin }.

%
%.  $Z_{min}=4$ and the horizontal axis shows the depth range, vertical axis shows the minimum layers $N$.

\section{EFS-based Dense-view Reconstruction} 
\label{sec:Dense Light Field Reconstruction Based on EFS}

Insufficient angular sampling causes aliasing in the focal stack, thus reconstructing a dense-view light field is equivalent to restoring a complete EFS corresponding to the non-aliasing light field/focal stack. As analyzed in Sec.\ref{sec:EFS_characteristics}, we can complement the non-aliased EFS by learning to recover the spectrum lines corresponding to missing viewpoints. Therefore, the light field reconstruction task is formulated as an EFS completion problem in this work. The pipeline of the proposed dense-view light field reconstruction is shown in Fig.\ref{fig:pipline}.

\begin{figure}
\vspace{-0.3cm}
\begin{center}
\centering
\includegraphics[width=0.95\linewidth]{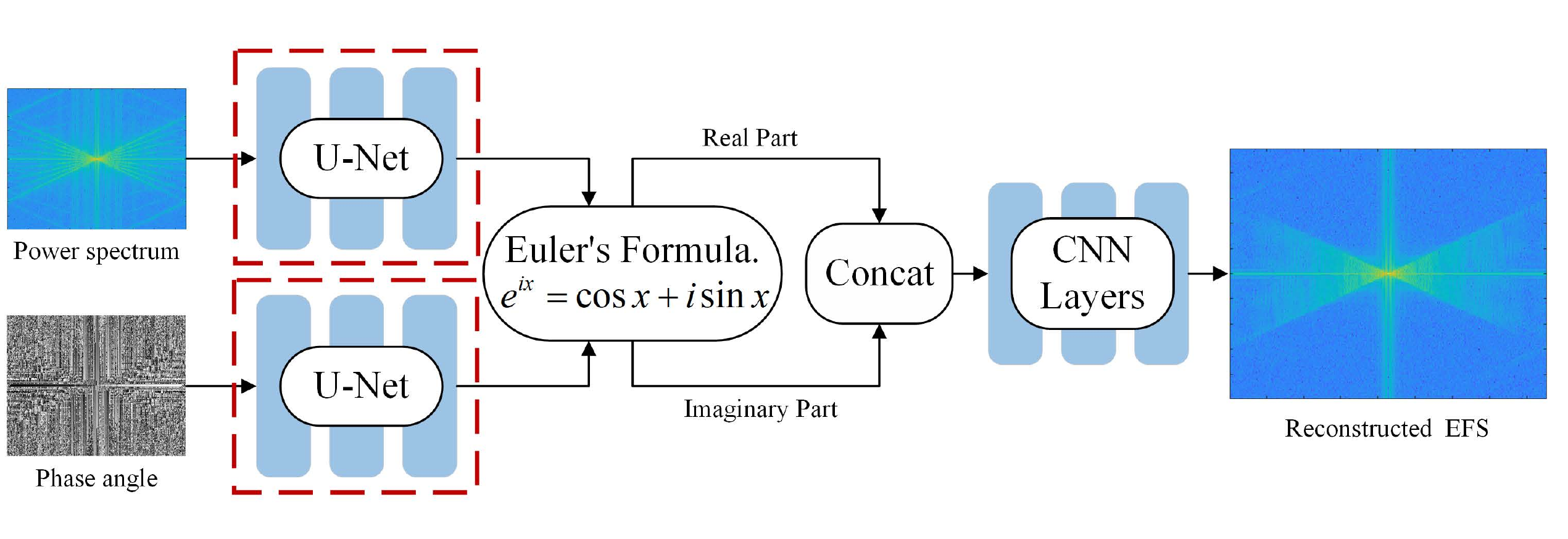}
\end{center}
\vspace{-0.3cm}
%\caption{(a) The dual-stream network architecture for non-aliasing EFS reconstruction. (b) Detailed architecture of the U-Net in (a).
\caption{The dual-stream network architecture for non-aliasing EFS reconstruction.}
\label{fig:network1}
\vspace{-0.6cm}
\end{figure}

%-------------------------------------------------------------------------
\subsection{EFS reconstruction} 
\label{sec:EFS reconstruction}
Specifically, We first perform shearing on an undersampled EPI to get the focal stack by Eq.\ref{eqn:focaltack}, then apply the Fourier transform on the aliased focal stack to get the aliased EFS (Eq.\ref{eqn:EFS_from_space}). Subsequently, a CNN $\phi$ parameterized by \textbf{$\sigma$} is proposed to complete the spectrum in the aliased $EFS_{ali}(\omega_f, \omega_x)$. The parameter $\sigma$ is optimized by

\begin{equation}
\small
  \label{eqn:EFS_reconstruct}
\underset{\sigma}{\arg \min }\left\{
\left\|{EFS}_{{gt}}, \phi_{\sigma}({EFS}_{ali})\right\|
\right\}.
\end{equation}

The loss function is,
\begin{equation}  \label{eqn:loss_symmetry}
\small
{ loss }=\Vert{EFS}-{{EFS}_{g t}}\Vert_{2}+\lambda{loss}_{s},
\end{equation}
where the scalar $\lambda$ is set to 1.5 for balancing the contributions of two loss terms. The second term ${loss}_{s}$ constrains the conjugate symmetry of the reconstructed EFS, 

\begin{equation} \label{eqn:loss_details}
 \small
{loss}_{s}=\frac{1}{N_{f} W} \sum\limits_{i=0}^{N_{f}-1} \sum\limits_{j=0}^{W-1}\left| {EFS}(\omega_{i}, \omega_{j})-  {{EFS}^*(-\omega_{i}, -\omega_{j})} \right|,
\end{equation}
%&{loss}_{s}=\frac{1}{C W} \sum\limits_{i=0}^{C-1} \sum\limits_{j=0}^{W-1}\left| |\mathcal{F}(\omega_{i}, \omega_{j})|-  |{\mathcal{F}(-\omega_{i}, -\omega_{j})}| \right|,
where $\!|\!\cdot\!|\!$ refers to the norm of a complex number and $*$ indicates the standard conjugate operation. %on a complex number. 
$N_f$ and $W$ are the number of refocus layers and the width of sub-aperture image respectively. 

As shown in Fig.\ref{fig:network1}, a novel dual-stream U-Net is designed to deal with complex number inputs. The power spectrum and phase angle are firstly fed into two sub-networks respectively. Then the features are combined using the Euler's formula to obtain the real and imaginary parts, which are concatenated and passed into several CNN layers for optimization. The detail of dual-stream U-Net is shown in the supplemental materials.
%Fig.\ref{fig:network1}(b) shows the detail of U-Net in Fig.\ref{fig:network1}(a). %where $N_f$ and  $W$ are the number of refocus layers and the image width respectively.
%Fig.\ref{network} shows the network architecture.

%\begin{figure}[!tb]
%\begin{center}
%\centering
%\includegraphics[width=\linewidth]{network_unet.pdf}
%\end{center}
%\caption{The detail of U-Net}
%\label{fig:network_unet}
%\end{figure}
%

\subsection{EPI reconstruction} 
\label{sec:EPI reconstruction}
The Fourier Slice Imaging Theorem \cite{ng2005fourier} tells a 2D slice through the origin of a 4D light field spectrum corresponds to a refocused image at a certain depth in the frequency domain. Based on this, we first apply the 1D inverse Fourier transform (IFT) of EFS along the $f$-axis, 
\begin{equation} \small \label{eqn:1D_fourier_inverse_transform}
\mathcal{F}(f,{\omega_x}) = \frac{1}{N_f}\sum\limits_{{\omega_f} = 0}^{N_f{ - 1}} {EFS({\omega_f},{\omega_x})} {e^{j2\pi \frac{{f{\omega_f}}}{N_f}}}.
\end{equation}

The projection relationship between $\mathcal{F}(f,{\omega_x})$ and $\mathcal{E}(\omega_u,\omega_x)$ can be obtained via Eq.\ref{eqn:fs_x_fourier}. Thus the Fourier spectrum  $\mathcal{E}_{efs}(\omega_u,\omega_x)$ of the reconstructed EPI %${E}_{efs}(u,x)$ 
could be calculated by performing a reverse projection,  %(Eq.\ref{eqn:inverse_fs_x_fourier}) to the Eq.\ref{eqn:fs_x_fourier}. After that, reconstructing $\mathcal{E}_{efs}(\omega_u,\omega_x)$ from $EFS(f,{\omega_x})$ is performed . 

% It is important to notice that  there is a transformation relationship between $f$ and $d$, as shown in Eq.\ref{shearing_f_d}
%\begin{equation}  \label{shearing_f_d}
%d=\frac{C+1}{2}-f, f=[1,C].
%\end{equation}

\begin{equation} \small \label{eqn:inverse_fs_x_fourier}
	\mathcal{E}_{efs}(\omega_u,\omega_x)=\mathcal{F}\left(-\frac{\omega_u}{\omega_x},\omega_x\right),
\end{equation}

\begin{figure}
\begin{center}
\centering
\includegraphics[width=0.9\linewidth]{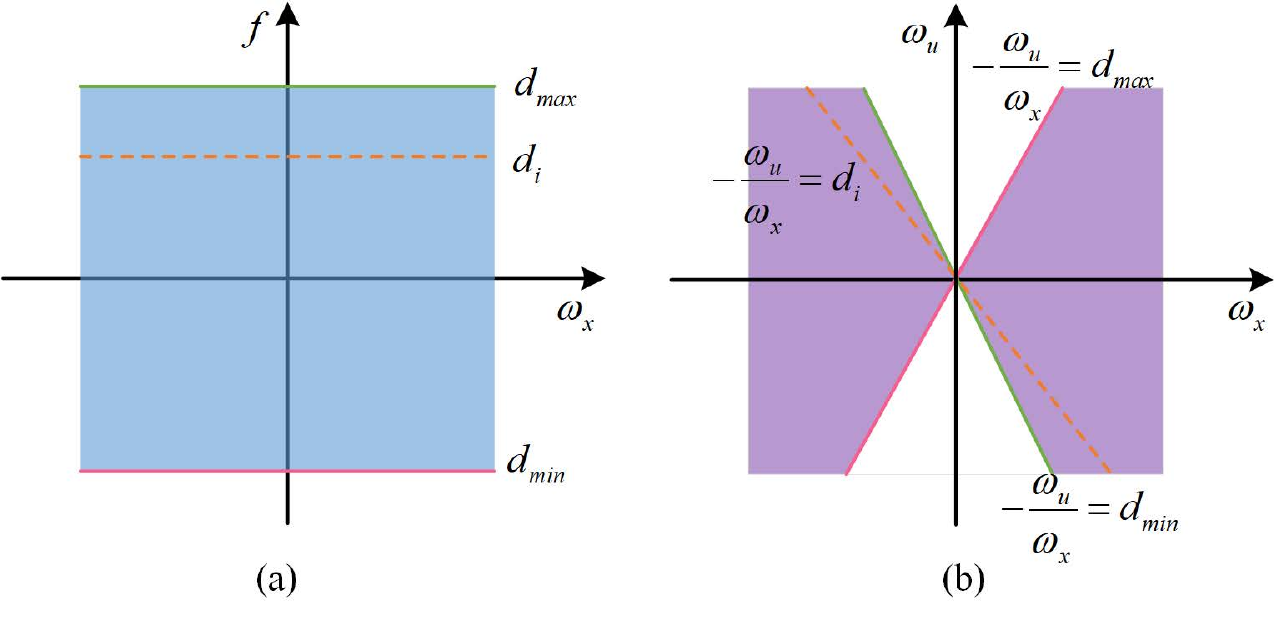}
\end{center}
\vspace{-0.3cm}
\caption{The diagram of the projection from Fig.\ref{fig:pipline}(d) to Fig.\ref{fig:pipline}(e). (a) The 1D IFT of EFS along the $f$-axis $\mathcal{F}(f,{\omega_x})$ (Fig.\ref{fig:pipline}(d)). (b) The reconstructed EPI spectrum $\mathcal{E}_{efs}(\omega_u,\omega_x)$ (Fig.\ref{fig:pipline}(e)). $[d_{min}$ $d_{max}]$ is the disparity (refocus) range of the scene.}  
\label{fig:efs_to_epi_proj.pdf}
\vspace{-0.5cm}
\end{figure}

Fig.\ref{fig:efs_to_epi_proj.pdf} shows the  diagram of this reverse  projection. Due to the limitation of disparity range, only the spectrum labeled as purple of  Fig.\ref{fig:efs_to_epi_proj.pdf}(b) can be reconstructed with this operation.

The next step is to apply the 2D IFT to get $E_{efs}(u,x)$ from $\mathcal{E}_{efs}(\omega_u,\omega_x)$. Since the interpolation operation is used during the focal stack construction, the ``tailing'' effects appear after the inverse Fourier slice operation, especially for the marginal viewpoints which are far away from the reference view. Hence we
%a u-net $\Psi_{\vartheta}$ with perceptual loss is used to optimize the reconstruction results.
use the U-Net $\Psi_{\mu}$ with a perceptual loss to optimize the reconstructed EPIs ${E}_{efs}$, 
\begin{equation} \small \label{eqn:EPI_reconstruct}
\underset{\mu}{\arg \min }\left\{
\left\|{E}_{{gt}}, \Psi_{\mu}({E}_{efs})\right\|
\right\},
\end{equation}
where ${E}_{gt}$ represents the ground truth EPI.

%\underset{\sigma}{\arg \min }\left\{
%\left\|{EFS}_{\mathrm{g}t}, \phi_{\sigma}({EFS}_a)\right\|
%\right\}.

%\underset{\mu}{\arg \min }\left\{\left\|{EPI}_{\mathrm{gt}},\Psi__{\mu}(EPI_{efs})\right\|\right\},

The loss function for optimization is defined as follow,
\begin{equation}  \label{eqn:EPI_loss}
loss_{EPI}=loss_{MAE}+ \gamma_{1}loss_{SSIM}+\gamma_{2}loss_{VGG},
\end{equation}
where $loss_{MAE}$ is the Mean Absolute Error loss, $loss_{SSIM}$ is the Structural Similarity loss \cite{wang2004image}, and $loss_{VGG}$ is the perceptual loss\cite{10.1007/978-3-319-46475-6_43} which is based on the VGG19 network trained on ImageNet. The scalars $ \gamma_i$ $(i=1,2)$ are set to 3 and 5 for balancing the effects of different loss terms. The detailed structure of this network is illustrated in the supplementary materials.

%\begin{figure}

%\begin{center}
%\centering
%\includegraphics[width=\linewidth]{network2.png}
%\end{center}
%\caption{Network architecture for optimizing the reconstructed EPI.}
%\label{fig:network2}
%\end{figure}

The complete dense-view light field reconstruction algorithm is given in Algorithm~\ref{algorithm1}. $H$ represents the height of the sub-aperture image.

\begin{algorithm}
\caption{}
\label{algorithm1}
\begin{algorithmic} [1]
\renewcommand{\algorithmicrequire}{\textbf{Input:}}
\renewcommand{\algorithmicensure}{\textbf{Output:}}
\REQUIRE ~~\\ An undersampled light field and disparity range $d_{range}$. The number of refocus layers $N_f$.

\ENSURE ~~\\ %算法的输出：Output
 	 Reconstructed dense-view light field.
\FOR {$i=1$ to $H$}
\STATE  Get the EPI $E(u,x)$.
\STATE  Perform the shearing operation on $E(u,x)$ within $d_{range}$ via Eq.\ref{eqn:epi_shearing}.\\
\STATE  Get the aliased $EFS_{ali}$ via Eq.\ref{eqn:focaltack}.\\
\STATE  Reconstruct the non-aliased ${EFS}(\omega_f,\omega_x)$ using the dual-stream U-Net  $\phi$ (as shown in Fig.\ref{fig:network1}).\\
\STATE  Perform 1D IFT on ${EFS}_{inv}(\omega_f,\omega_x)$ via Eq.\ref{eqn:1D_fourier_inverse_transform}. \\
\STATE  Reconstruct $\mathcal{E}_{efs}(\omega_u,\omega_x)$ via Eq.\ref{eqn:inverse_fs_x_fourier}.\\

\STATE  Perform 2D IFT on $\mathcal{E}_{efs}(\omega_u,\omega_x)$ to get $E_{efs}$.\\

\STATE   Optimize the reconstructed EPI with the U-Net $\Psi$.
\ENDFOR
\STATE Output the reconstructed dense-view light field. 
\end{algorithmic}
\end{algorithm}
%-------------------------------------------------------------------------
\section{Evaluations}
\label{sec:Evaluations}
To evaluate our proposed EFS-based dense-view light field reconstruction method, we conduct experiments on both synthetic and real light field datasets. The real light field datasets are captured by both the camera array and the plenoptic camera (Lytro Illum \cite{lytro_web}). We mainly compare our approach with two state-of-the-art learning-based methods, Wu {\it {et al.}} \cite{wu2018light} (without explicit depth estimation) and LLFF \cite{mildenhall2019local} (MPI-based). Note that LLFF \cite{mildenhall2019local} is retrained on our training date using the released training code for fair comparison. For Wu {\it {et al.}} \cite{wu2018light} without the original code being released, we just use the trained model provided by the authors. 

To empirically validate the robustness of the proposed method, we perform evaluations under different sampling patterns. The quantitative evaluations are performed by measuring the average PSNR and SSIM metrics over the synthetic views of the luminance channel. We also analyse the spectrum energy losses for EFS reconstruction and EPI reconstruction respectively. 

% Comparative experiments with STOAs are also provided to show the superiority of proposed method.
%-----------------------------------------------------------------------------

\subsection{Datasets and Implementation Details }
\label{sec:Datasets and Implementation Details}
In the training process, both the synthetic and real light fields are used. For the synthetic data, 12 light fields containing complex textured structures are rendered using the automatic light field generator \cite{zhu2019tvcg,povray_web}, of which 7 are for training and 5 for testing. Real light fields are taken from the high-resolution light field dataset provided by Guo {\it {et al.}} \cite{guo2018dense}, of which 20 are for training and 6 for testing. In order to show the relationship between viewpoints and EFS lines, we utilize the first 200 viewpoints for experiments. Additionally, the light fields from the Disney \cite{kim2013scene} dataset are used to verify the performance of the proposed method on unseen scenes captured by a camera array. Tab.\ref{tab:data parameters} displays the parameters of all the datasets. In the dense-view light fields, the disparity between two adjacent views for most scenes is less than one pixel, while in several scenarios, the disparity reaches two pixels.

\begin{table}[t]
\footnotesize
\begin{center}
\caption{Parameters of the light field (LF) datasets.}
\vspace{-0.2cm}
\label{tab:data parameters}
%\begin{tabular} { lcc }
\begin{tabularx}{0.4\textwidth}{@{} lCC @{}}
\hline
\hline
Dataset & Angular Res. & Spatial Res.\\
\hline
Syn. LFs  & $ 1\times 200 $ & $512\times 512$  \\ 
Real LFs \cite{guo2018dense} & $ 1\times 200 $ & $376\times 512$\\ 
Couch \cite{kim2013scene} & $ 1\times 101 $ &  $628\times 1024 $\\
Church \cite{kim2013scene} & $1\times 101 $ &  $670\times 1024$\\
%StillLife\cite{hci_web}&[-1,0.98] & $1\times 9$  & $768\times 768$\\
%\hline
Bike\cite{kim2013scene} & $1\times 51 $ &  $670\times 1024$\\ 
Statue\cite{kim2013scene} & $1\times 151 $ &  $670\times 1024$\\ \hline
\hline
%Lego \cite{stanford_data}&[-1.00,0.98] & $1\times 17 $ & $1024\times 1024$\\ \hline
\end{tabularx}
%\end{tabular}
\end{center}
\vspace{-0.5cm}
\end{table}

At present, only the disparity along one single direction is concerned so that the 2D EPIs can represent the input light field. The details of several light fields under 10$\times$ and 15$\times$ downsampling scales is shown in the supplementary materials. For each light field in the experiment, by considering the disparity and the scene distribution, the EFS is constructed by performing the refocus operation for 200 times ($N_f=200$) where $\Delta \alpha=0.01$. The range of scene disparity determines the range of refocusing operation $f$. For details regarding the disparity, refer to our datasets and codes released later.

% The resolutions of real scenes and synthetic scenes are $121\times376\times526 $ and $121\times526\times526$ respectively.
%\textbf{Implementation Details.}
%In order to prove that the EFS is capable of dealing with large disparities, we conduct experiments with different angular sampling rates. At present, only horizontal disparities are concerned, so the 2D EPI image can be used to represent the input Light field. We consider 10 times downsampling and 15 times downsampling respectively. The specific sampling pattern is given in Fig.\ref{differ_samlpy_of_efs}

%\begin{figure}
%    \begin{center}
%    \includegraphics[width=\linewidth]{differ_epi_samply.pdf}
%    \end{center}
%    \caption[example]
%    {EPIs under different viewpoint downsampling rates \label{fig:differ_epi_samply}. The disparity range of each EPI is labelled on the image. (a) One viewpoint image of a light field. (b) Original EPI with dense viewpoint sampling. (c) 10$\times$ downsampling of (b). (d) 15$\times$ downsampling of (b). 
%    }
%\end{figure}

\subsection{Spectrum Domain}
\label{sec:Spectrum Domain}
\begin{figure}
\begin{center}
\centering
\includegraphics[width=\linewidth]{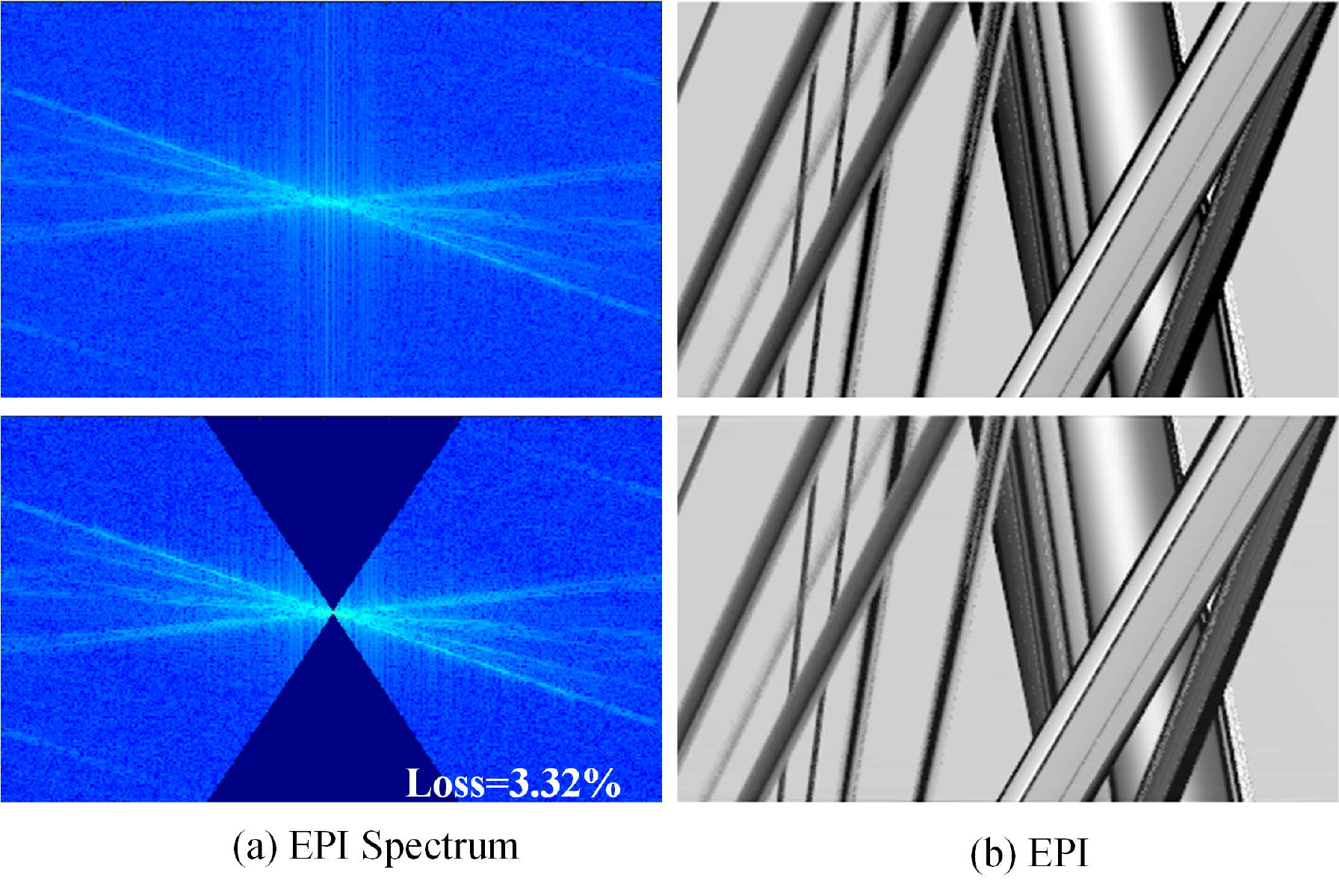}
\end{center}
\vspace{-0.4cm}
\caption{An example of the original EPI spectrum and the EPI spectrum projected from the EFS. The upper row displays the whole dense EPI spectrum from the ground truth EPI. The lower row displays the EPI spectrum obtained within a scene disparity range [-1,1], which has a 3.32$\%$ loss, and the reconstructed EPI.}
\label{fig:one_epi_loss.pdf}
\vspace{-0.5cm}
\end{figure}

\indent \textbf{EPI spectrum loss analysis}. In our experiments, the disparity range $d_{range}=[d_{min},d_{max}]$ is also the focal stack range. %In other words, \textcolor{red}{each 3D point shows the entity in the focus image in the focal stack}. 
With the infinite aperture assumption, the spectrum energy could be regarded as zero for other focus layers beyond the constructed focal stack\cite{levin2010linear}. However, since the infinite aperture camera is currently not available, the EFS representation for a light field has a certain loss. Fig.\ref{fig:one_epi_loss.pdf} shows the difference between the whole dense-view EPI spectrum and the EPI spectrum within a certain disparity range obtained from the EFS. Although partial contents are missing in the latter case (about 3.32$\%$), the structural information on the EPI can still be reconstructed. 

\begin{figure}
\begin{center}
\centering
\includegraphics[width=\linewidth]{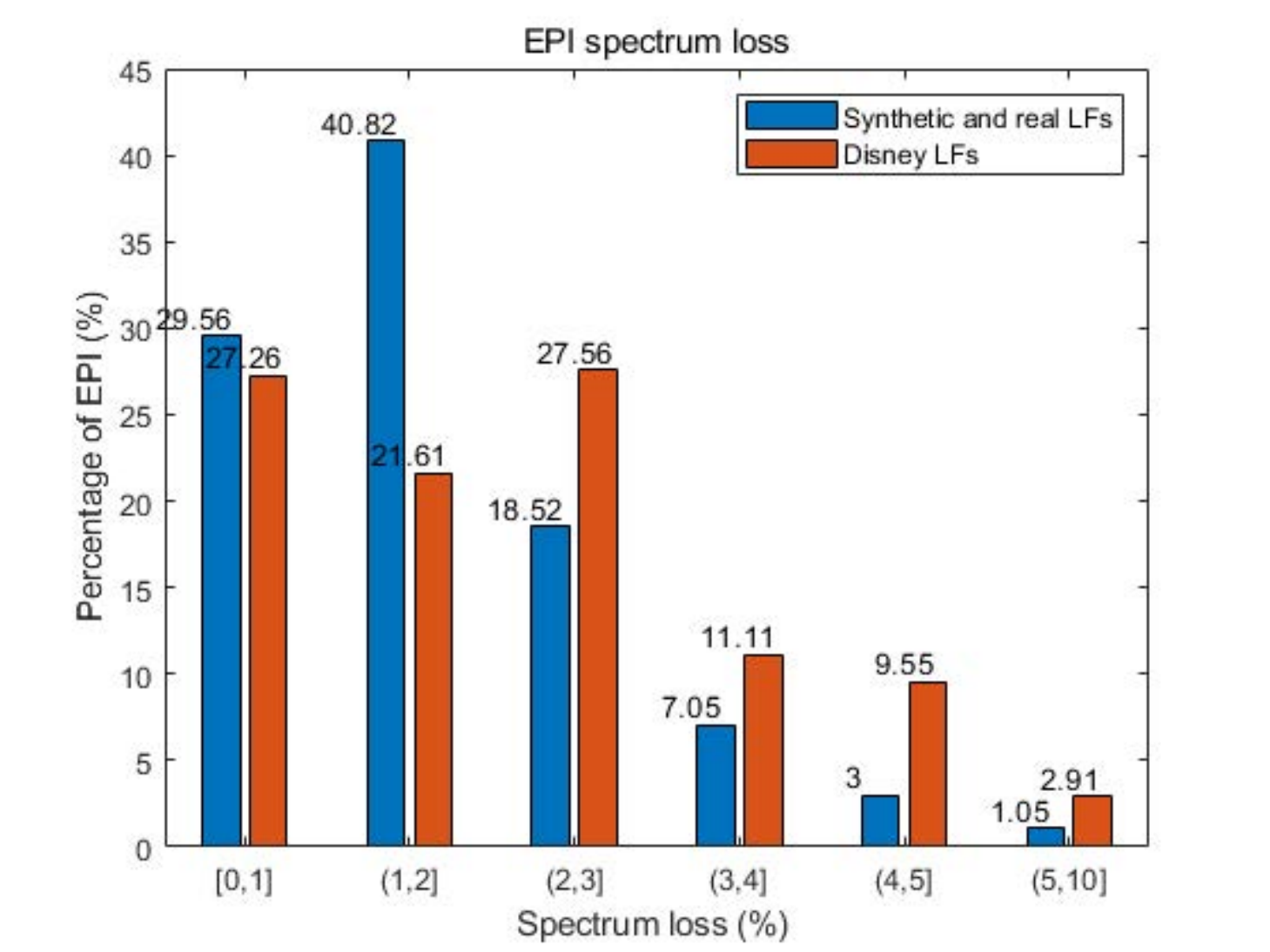}
\end{center}
\vspace{-0.4cm}
\caption{Distributions of EPI spectrum loss on several light field datasets.}
\label{fig:epiloss}
\vspace{-0.5cm}
\end{figure}

To evaluate the upper bound of the EPI spectrum loss in a dense-view light field, we have counted the spectral energy loss over 10,000 EPIs which are reconstructed from the EFS containing all the scene depth ranges for several light fields and summarized the results in Fig.\ref{fig:epiloss}. In these EPI spectra, the maximum loss is 9.03$\%$, the minimum loss is 0.31$\%$, and the average loss is 1.94$\%$. It can be seen that the EPI spectrum loss generally has a sparse ($\leq$ 5$\%$) distribution.
%It can be seen that the EPI spectrum loss is generally lower than 5$\%$. 
In addition, the distribution of the loss on Disney LFs\cite{kim2013scene} is flatter and more spread out than that on synthetic\cite{povray_web} and real LFs\cite{guo2018dense}. 
%The loss distribution on Disney data is more biased towards a high loss range.
We attribute this to: 1) the background texture of Synthetic LFs is relatively simple, while the texture of Disney LFs is more complex, 2) the scene depth of Synthetic data is primarily concentrated within a 
limited interval, while the depth distribution of Disney LFs is more divergent.

% The loss distribution on Disney data is more biased towards a high loss range.
%此外，从表中我们还可以看出，真实场景或者相机阵列拍摄的（diseny）场景loss 更高，这是因为真实场景纹理更复杂，深度分布更发散。而仿真数据其深度范围大多集中在某一特定区间，背景纹理也较为简单。

\begin{figure}[t]
\begin{center}
\centering
\includegraphics[width=\linewidth]{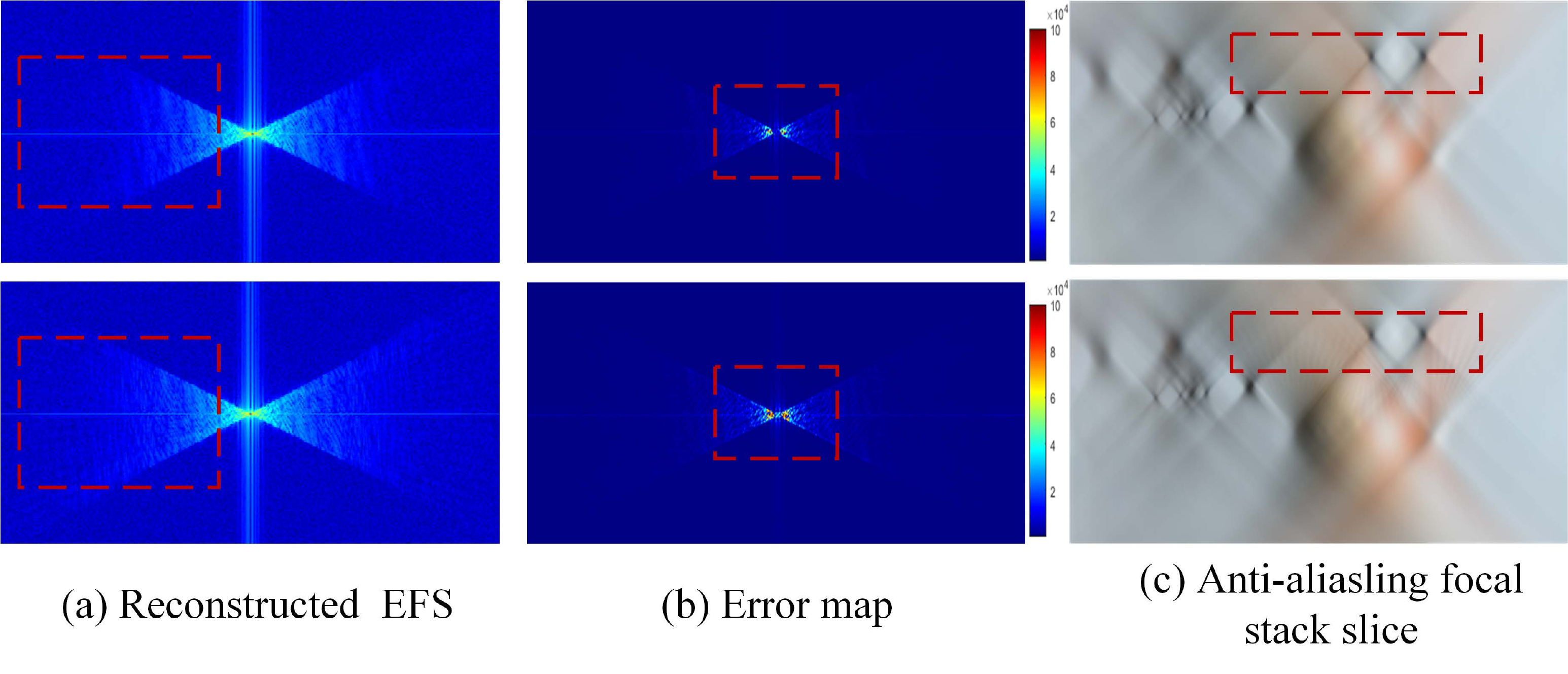}
\end{center}
\vspace{-0.4cm}
\caption{EFS reconstruction results under different  downsampling settings. From top to bottom: 10$\times$ and 15$\times$ downsampling. (a) Reconstructed EFS. (b) Error
map of the EFS. (d) Corresponding anti-aliasing focal stack.}
\label{fig:errorofefs}
\end{figure}

\textbf{EFS reconstruction loss analysis}. We try to analyze the loss in the EFS reconstruction. Some results are displayed in Fig.\ref{fig:errorofefs}, which consists of reconstructed EFS, the error map, and corresponding anti-aliasing focal stack.

%and Tab.{??} shows the average EFS energy loss on different test scenarios at different sampling rates,

By comparing Fig.\ref{fig:differ_samlpy_of_efs} (b)(c) with Fig.\ref{fig:errorofefs}, our method achieves a preferable performance on the EFS reconstruction for different sampling rates. The average EFS energy loss with different sampling rates on several test scenarios is list in Tab.\ref{tab:EFS energy loss}.

\begin{table}[t]
\begin{center}
\caption{Average EFS spectral energy loss on different test scenarios.}
\vspace{-0.2cm}
\label{tab:EFS energy loss}
\begin{tabular}{  p{2cm}<{\centering}p{2cm}<{\centering}p{2cm}<{\centering} }
\hline
\hline
Dataset & 10$\times$  & 15$\times$  \\
\hline
Syn. LFs &2.74\% & 3.53\%\\
Real LFs \cite{guo2018dense} &2.78\%& 3.56\%\\ 
Disney LFs \cite{kim2013scene}& 1.39\%  & -\\
\hline
\hline
\end{tabular}
\end{center}
\vspace{-0.4cm}
\end{table}

%\textcolor{red}{(not suitable for this sub-section, maybe for 5.3, also double check the table index used here)} By comparing the last column of Tab.\ref{tab:All Quantitative results} with Tab.\ref{tab:EFS energy loss}, we find that the average EFS spectral energy loss has a significant influence on PSNR while having little effect on SSIM \textcolor{red}{(how to get this conclusion?)}. This demonstrates the robustness of the proposed reconstruction method, considering the fact that human eyes are more sensitive to changes of the structure information (SSIM) than individual pixel values (PSNR) \cite{ssim_paper}.
%This feature is important especially for refocused images where the errors in defocus blur are hard to be distinguished by people.  So, a higher SSIM in Tab.\ref{tab:average of quantitative results} could better demonstrate the advantages of the proposed method.

\subsection{Parameter analysis}
\label{sec:Ablation Studies}
\begin{figure*}
\begin{center}
\centering
\includegraphics[width=\linewidth]{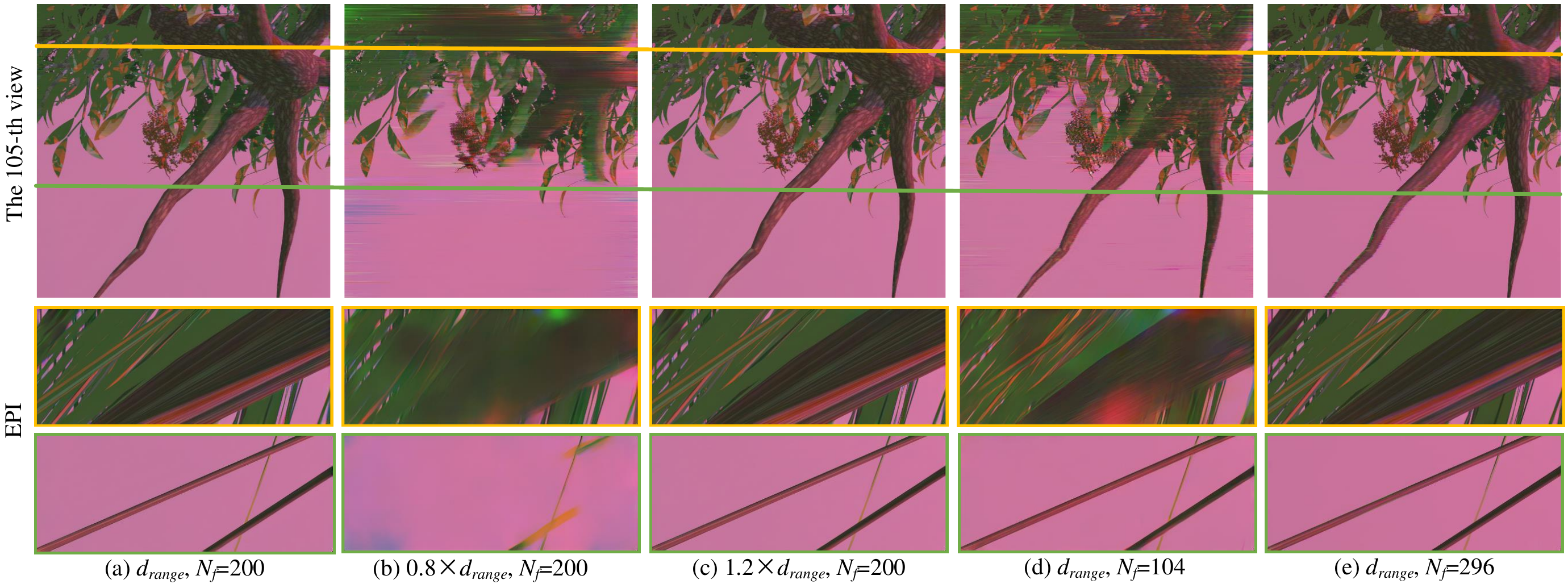}
\end{center}
\vspace{-0.4cm}
\caption{Qualitative comparisons regarding the range of refocus operation and the number of EFS layers on a synthetic light field. The top row shows reconstructed views with different parameters. The remaining two rows show reconstructed EPIs  corresponding to the yellow and green lines in the reconstructed view respectively. %(a),(b) and (c) have the same  $N_f=200$, but the range of refocus operations are different. (a),(d) and (e) have the same range of  refocus operations  $d_{range}$, however, the  number of EFS layers $N_f$ are different.(f) Ground truth
}
\label{fig:Ablation}
\end{figure*}

We empirically validate the influences of the range of refocus operation and the EFS sampling on the reconstructed EPI by performing the following parameter analysis. In these experiments, we use our synthetic light fields with 15$\times$ downsampling.

\textbf{The range of refocus operation}. The refocus range is set to 0.8$d_{range}$, $d_{range}$ and 1.2$d_{range}$, respectively. As mentioned in Sec.\ref{sec:Spectrum Domain},  with 15$\times$ downsampling of the original light field, $d_{range}$ is the scene disparity range ($[15d_{min},15d_{max}]$). Fig.\ref{fig:Ablation} (a)(b)(c) show the qualitative comparisons with different refocus ranges on the tree root dataset. Quantitative analysis, in terms of average PSNR and SSIM, is summarized in the 2nd, 3rd and 4th rows of Tab.\ref{tab:Ablation}. As shown in Fig.\ref{fig:Ablation} (b), partial tree root has not been reconstructed. The scene structure can not be reconstructed completely when the range of refocus operation is too narrow.

\textbf{The number of EFS layers $N_f$}. The number of refocus layers $N_f$ is set to 104, 200 and 296, respectively. Fig.\ref{fig:Ablation} (a)(d)(e) show the qualitative comparisons with different numbers of EFS layers. The 2nd, 5th, and 6th rows of Tab.\ref{tab:Ablation} show the quantitative comparison on reconstructed light fields. It is noticed that insufficient focal layers, of which the count is smaller than the minimum focal layer count $N_{fmin}$ (see Sec.\ref{sec:Sampling analysis of EFS}), \textit{i.e.}, $N_f$=104, will cause performance degradation ($32.92/0.823$ ${vs}$ $37.54/0.952$ in terms of PSNR and SSIM respectively). In this case, aliasing appears in the focal stack, which leads to over-smooth textures in Fig.\ref{fig:Ablation}(d). In addition, when $N_f$ is increased from 200 to 296, only a slight improvement in the performance is reported ($37.54/0.952$ ${vs}$ $37.98/0.961$). % in terms of PSNR and SSIM respectively).
Hence once the number of EFS layers $N_f$ meets the minimum sampling rate requirement, continuously increasing EFS layers would not bring obvious improvements in the performance.

%Combined with Fig.\ref{fig:Ablation} and  Tab.\ref{tab:Ablation}, we can see that after EFS layers reach the number of $N_{fmin}$ analyzed in section\ref{sec:Sampling analysis of EFS}, increasing the number of EFS layers does not have great influence on the reconstruction quality, however, once the sampling of EFS layers is not enough, the image reconstruction quality will be greatly reduced.

\begin{table}[t]
\vspace{-0.2cm}
\begin{center}
\caption{Average PSNR and SSIM of Fig.\ref{fig:Ablation}.}
\vspace{-0.2cm}
\label{tab:Ablation}
\begin{tabular} {cccc} 
\hline\hline
Refocus range $d_{range}$ & \# EFSs $N_f$& PSNR$\uparrow$& SSIM$\uparrow$ \\
\hline
$0.8\times$ & 200 & 35.05 & 0.865\\
$1.0\times$ & 200 & 37.54 & 0.952\\
$1.2\times$ & 200 & 37.96 & 0.959\\
$1.0\times$ & 104 & 32.92 & 0.823\\
$1.0\times$ & 296 & 37.98 & 0.961\\ \hline\hline
\end{tabular}
\end{center}
\vspace{-0.5cm}
\end{table}
%\begin{table}[t]
%	\small
%	\begin{center} 
%		\caption{Average of spectral energy loss on different  test scenarios.}
%		\label{tab：spectral energy loss}
%\begin{tabular}{lcc}
%\hline
%Data& 5$\times$  & 15$\times$  \\
%\hline
%synthetic Light field&2.03\% & 3.58\%\\ 
%Guo $et~al.$\cite{guo2018dense} &2.50 \%& 3.13\%\\
%Disney Light field\cite{kim2013scene}&  \multicolumn{2}{c}{0.665\% (10$\times$ )} \\
%\hline
%\end{tabular}
%	\end{center}
%	\vspace{-1.8em}
%\end{table}
%

\subsection{Comparisons with SOTAs}
\label{sec:Comparison with SOTAs}
We compare our method against Wu {\it {et al.}} \cite{wu2018light} and LLFF \cite{mildenhall2019local}. Tab.\ref{tab:All Quantitative results} shows the average PSNR/SSIM/LPIPS\cite{zhang2018unreasonable} measurements with different downsampling rates on both synthetic and real light fields. Qualitative comparisons between different methods on several test scenes under 15$\times$ downsampling rate are shown in Fig.\ref{fig:sota_syn}, Fig.\ref{fig:sota_realdata} and Fig.\ref{fig:sota_camera_array} respectively.

%In this section, we evaluate the proposed approach using synthetic light field datasets, real light field datasets and camera array datasets, respectively.  Fig.\ref{?},Fig.\ref{?} and Fig.\ref{?} are the comparison results of different methods under 15$\tims$ downsampling rates 

\subsubsection{Synthetic light field datasets}
\begin{figure*}
\vspace{-0.4cm}
\begin{center}
\centering
\includegraphics[width=\linewidth]{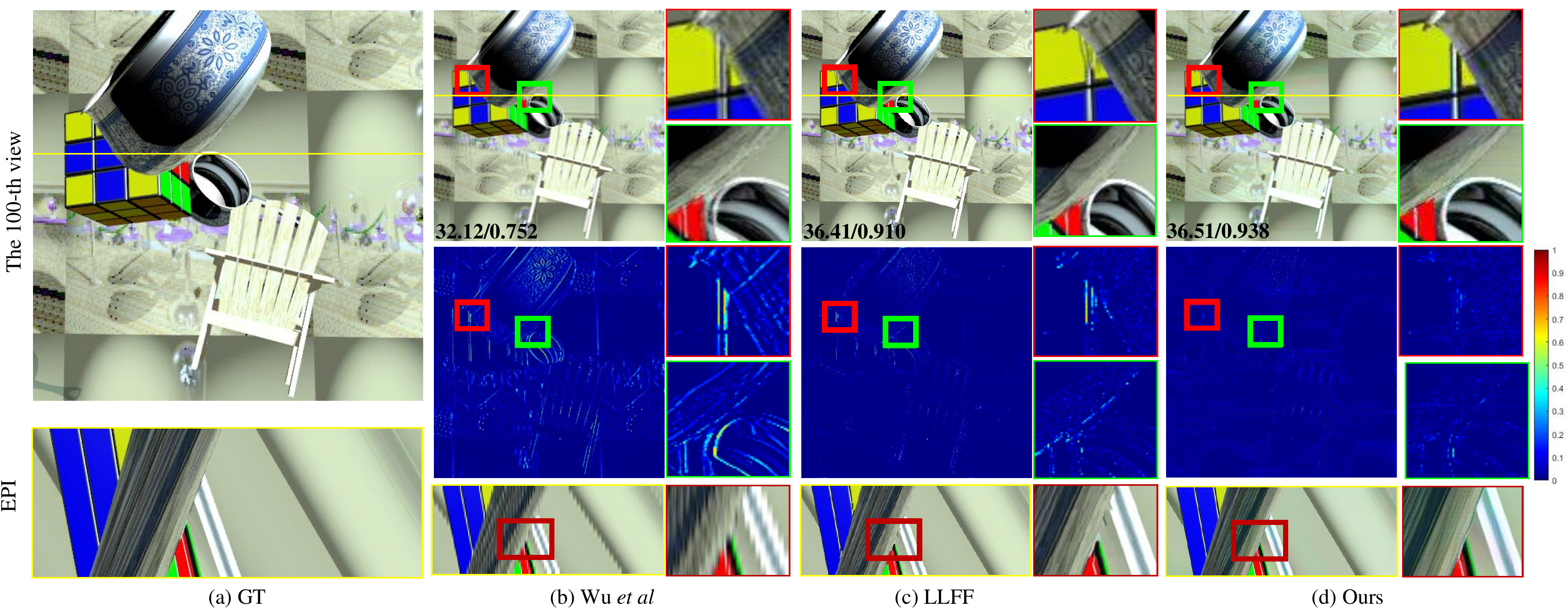}
\end{center}
\vspace{-0.4cm}
\caption{Comparison of reconstruction results on the synthetic light field dataset (15$\times$ downsampling). The results include one reconstructed view (with the PSRN and SSIM measurements overlaid), the error map, and the EPI of the reconstructed light field by different methods. Several local areas are zoomed in for better visualization. From left to right: (a) GT, results by (b) Wu {\it {et al.}} \cite{wu2018light}, (c) LLFF \cite{mildenhall2019local}, and (d) our method.} 
\label{fig:sota_syn}

\end{figure*}

We evaluate the proposed approach using our synthetic light field datasets under 10$\times$ and 15$\times$ downsampling rates. Qualitative results under 15$\times$ downsampling (maximum disparity up to 8px) are shown in Fig.\ref {fig:sota_syn}. 

As shown in Fig.\ref{fig:sota_syn}(b), ghosting artifacts are visible around the boundary region in the reconstruction result by Wu {\it {et al.}} \cite{wu2018light}, which are caused by the limited receptive field of their network. Also, the Gaussian convolution kernel is only effective for small disparity. The MPI-based LLFF \cite{mildenhall2019local} tends to assign high opacity to incorrect layer for the region with ambiguous/repetitive texture or moving content between input images, which will cause ﬂoating or blurred patches around the boundary region (see the boundary region of the pot in Fig.\ref{fig:sota_syn}(c)). In comparison, the proposed EFS-based reconstruction method produces more clear boundaries (as shown in Fig.\ref{fig:sota_syn}(d)).  

Fig.\ref{fig:sota_syn_psnr_ssim} shows the PSNR and SSIM measurements for each reconstructed view on the synthetic light field under 10$\times$ and 15$\times$ downsampling rates. Due to the shearing process used in our method (Eq.(\ref{eqn:epi_shearing})), the more marginal views there are, the more image information will be sheared out of the image. Thus the reconstruction results are not satisfactory on these marginal views. Still, the overall performance of our method is better than the SOTAs, especially for larger disparity scenarios. The 3rd and 4th columns of Tab.\ref{tab:All Quantitative results} list the quantitative measurements on synthetic light fields under 
different downsampling rates, which further validates the superiority of the proposed method. 
% What is more,  reconstructed light fields with higher view consistency  (as shown in the demonstrated EPIs)越是边缘的视点，就会有越多图像信息被shear到图像外面去了，所以这块其实是有损失的,重建效果就不太理想，但是我们的方法总体上优于3 和2，而且视差越大效果越明显。为了消除这一问题，可以先对光场空间维度进行padding,对padding后的光场进行重建.fu

%由于llff对场景的划分粒度   对于视差范围大的，深度划分(MPI)不够稠密，导致图像边缘出现
%（？？），但是当增加MPI 层数以后，所需要的计算代价就是增大。

\subsubsection{Real light fields captured with a plenoptic camera}
\begin{figure*}
\begin{center}
\centering
\includegraphics[width=\linewidth]{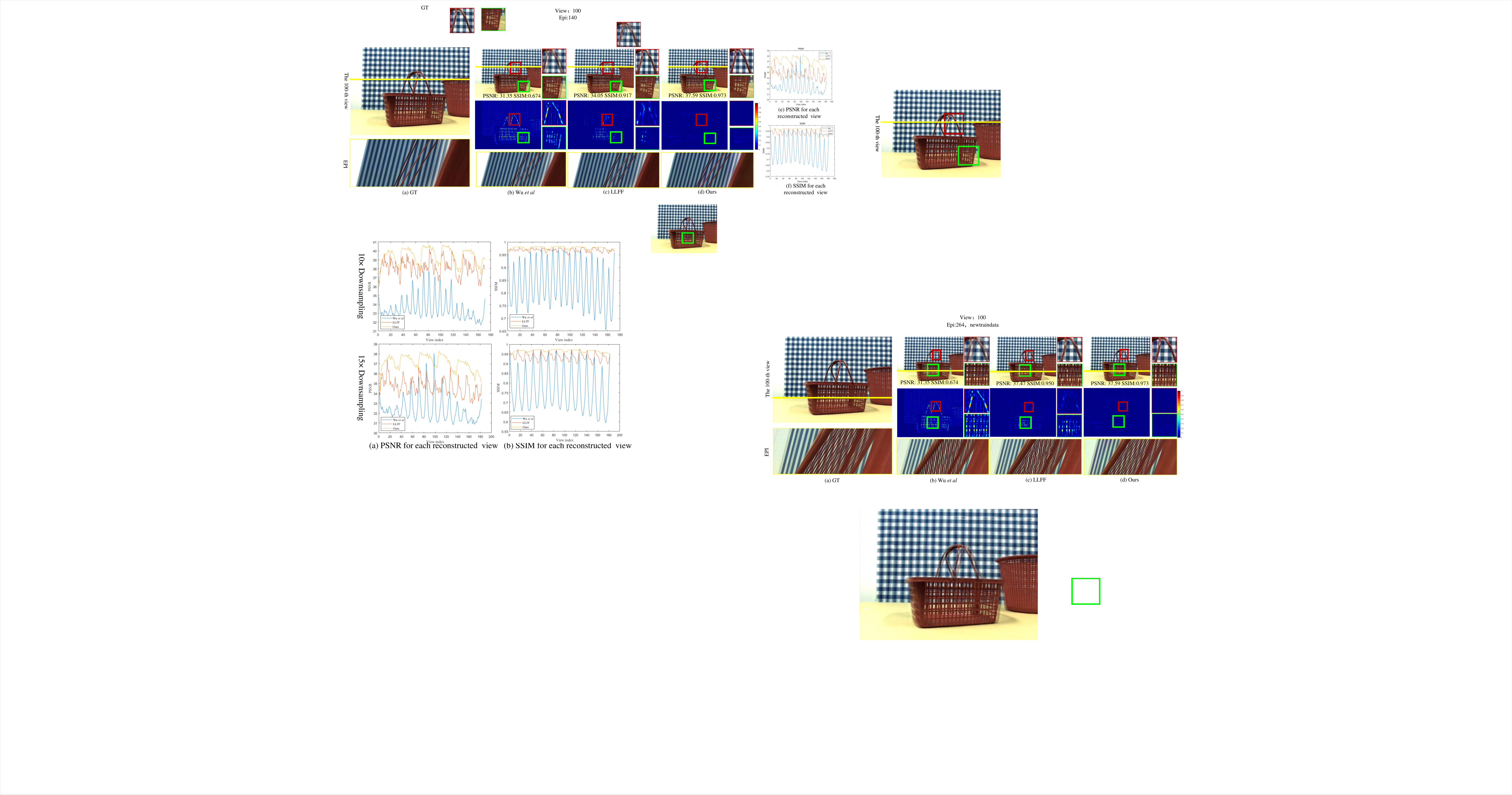}
\end{center}
\vspace{-0.4cm}
\caption{Comparison of reconstruction results on the real light field dataset (15$\times$ downsampling). The results include one reconstructed view, the error map, and the EPI of the reconstructed light field by different methods. Several local areas are zoomed in for better visualization. From left to right: (a) GT, results by (b) Wu {\it {et al.}} \cite{wu2018light}, (c) LLFF \cite{mildenhall2019local}, and (d) our method.}
\label{fig:sota_realdata}
\vspace{-0.4cm}
\end{figure*}

We also evaluate the proposed approach using the Lytro light field dataset \cite{guo2018dense}, which contains massive static scenes in the real world, such as bicycles, toys, and plants. These scenes are challenging in terms of abundant colors and complicated occlusions. 

Fig.\ref{fig:sota_realdata} shows the reconstruction results of ``basket'' under 15$\times$ downsampling. There exist many thin structures in the scene, such as the basket handle. The texture on such a thin structure changes very fast, which results in difficulties for reconstruction. We can see that severe ghosting artifacts occur around the basket handle in the results by Wu {\it {et al.}} \cite{wu2018light} and reconstructed views are inconsistent (Fig.\ref{fig:sota_realdata}(b)). Similarly, a fuzzy phenomenon appears in the results by LLFF \cite{mildenhall2019local} on the handle part. By reconstructing the dense light field in the frequency domain, our method is less sensitive to the spatial contents, and thus capable of producing high-quality and consistent view reconstruction.

\begin{figure}
\begin{center}
\centering
\includegraphics[width=\linewidth]{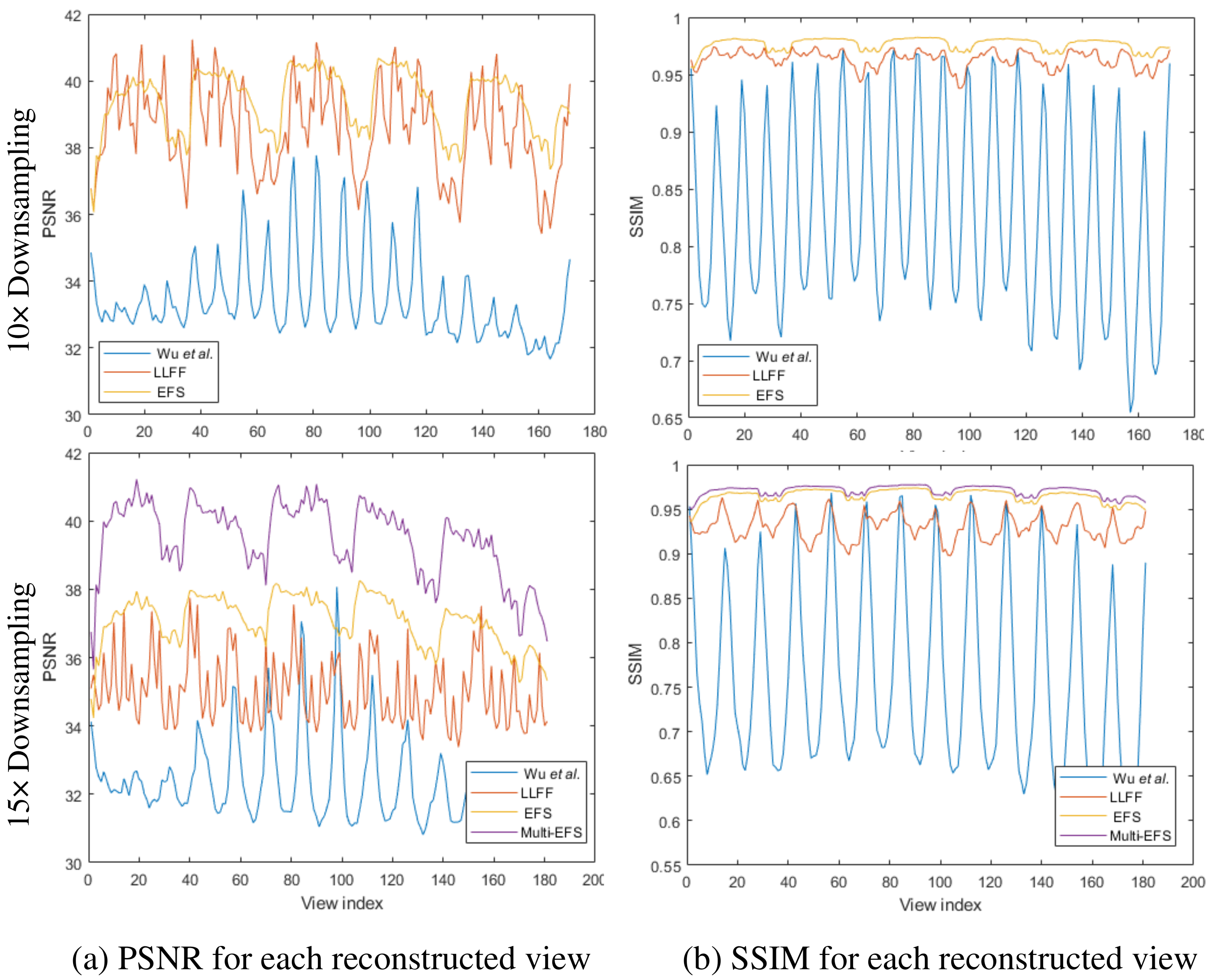}
\end{center}
\caption{PSNR and SSIM for each reconstructed view of Fig.\ref{fig:sota_realdata} under 10$\times$ and 15$\times$ downsampling.}
\label{fig:sota_real_psnr_ssim}
\end{figure}

Fig.\ref{fig:sota_real_psnr_ssim} shows the PSNR and SSIM measurements for each reconstructed view of Fig.\ref{fig:sota_realdata} under 10$\times$ and 15$\times$ downsampling. As for the high complexity of the scene, over 95$\%$ view reconstruction results by our method are better than SOAT methods. Quantitative comparisons in terms of PSNR, SSIM, and LPIPS are listed in the 5th and 6th columns of Tab.\ref{tab:All Quantitative results}.

\subsubsection{Real light fields captured with a camera array}

\begin{figure*}
\begin{center}
\centering
\includegraphics[width=\linewidth]{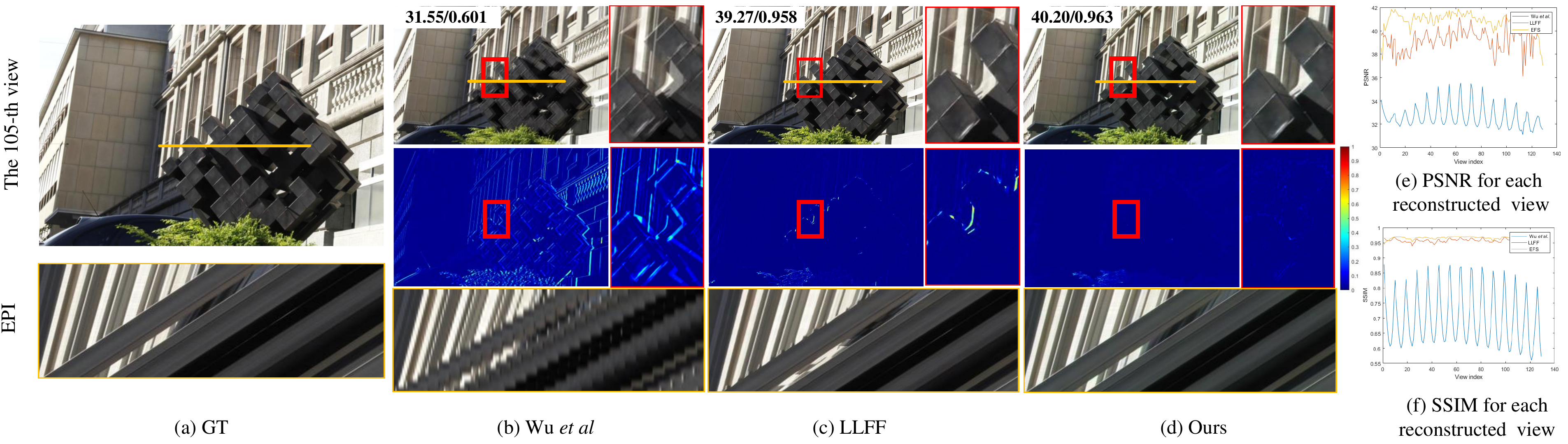}
\end{center}
\vspace{-0.4cm}
\caption{Comparison of reconstruction results on the camera array light field dataset (15$\times$ downsampling). The results include one reconstructed view, the error map, and the EPI of the reconstructed light field by different methods. Several local areas are zoomed in for better visualization. From left to right: (a) GT, results by (b) Wu {\it {et al.}} \cite{wu2018light}, (c) LLFF \cite{mildenhall2019local}, and (d) our method, quantitative results in terms of (e) PSNR and (f) SSIM.}
\label{fig:sota_camera_array} %同样可以验证泛化性
\end{figure*}

\begin{figure}
\begin{center}
\centering
\includegraphics[width=\linewidth]{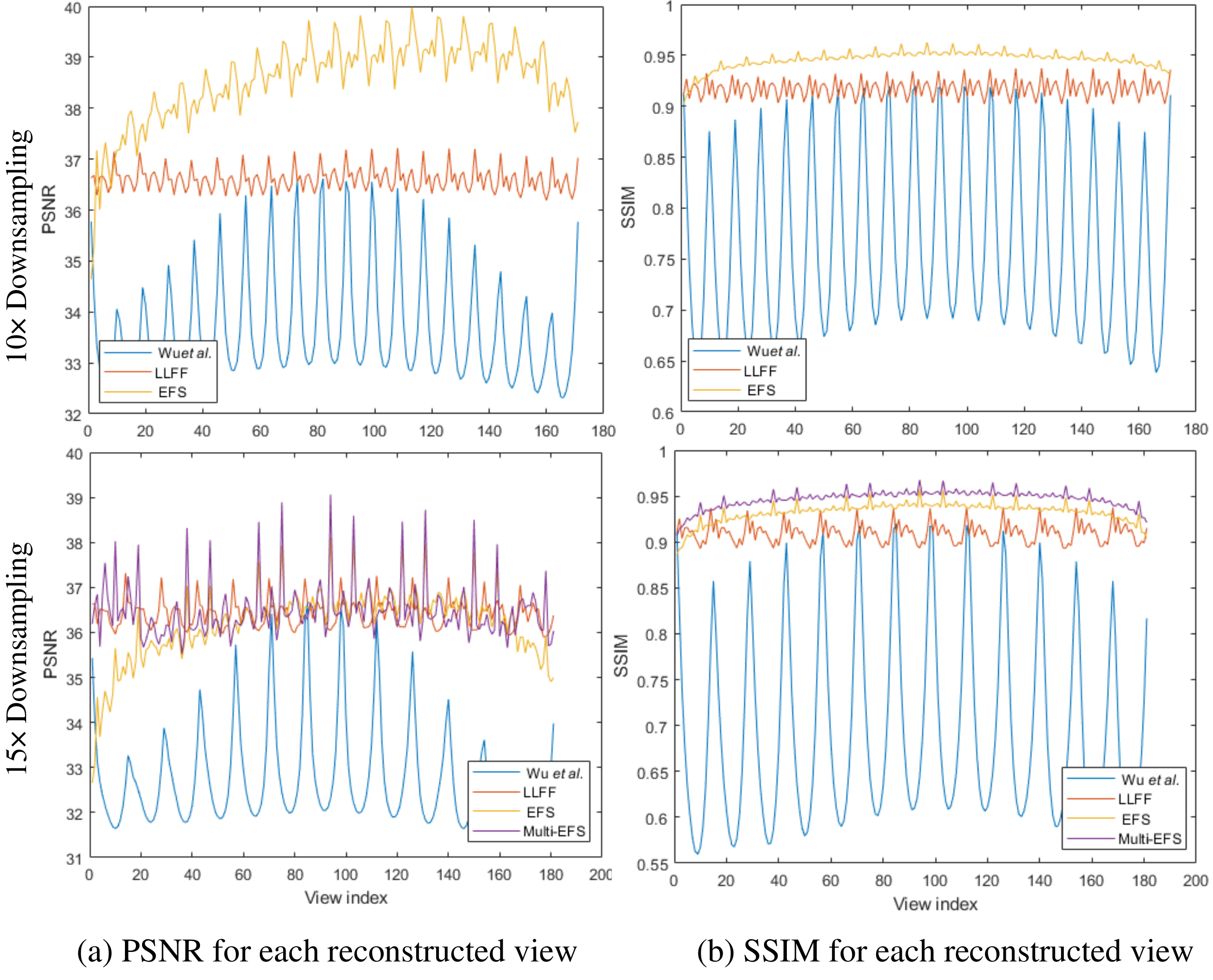}
\end{center}
\caption{ PSNR and SSIM for each reconstructed view of Fig.\ref{fig:sota_syn} under 10$\times$ and 15$\times$ downsampling.}
\label{fig:sota_syn_psnr_ssim}
\end{figure}

In order to verify the effectiveness of our method under wide baseline and large disparity conditions, we further evaluate the proposed approach using the Disney light fields \cite{kim2013scene} which are captured by a camera array.

Fig.\ref{fig:sota_camera_array} shows the reconstruction results on the statue scene under 10$\times$ downsampling (maximum disparity up to 15px). Due to the limited receptive field of the networks, the results by Wu {\it {et al.}} \cite{wu2018light} show serious aliasing effects on all the foreground objects (Fig.\ref{fig:sota_camera_array}(b)). Due to the memory limitation, there is a trade-off between the image resolution and the layers of MPIs utilized by LLFF\cite{mildenhall2019local}, which leads to performance degradation for large disparity areas with high-resolution input. Also severe artifacts appears in the regions with repetitive patterns, large disparity and occlusions, as shown in the zoom-in rectangles in Fig.\ref{fig:sota_camera_array}(c). 
%the MPI network sometimes assigns high opacity to incorrect layers in ambiguous or repetitive texture regions, especially in occlusion boundary areas. In Fig.\ref{fig:sota_camera_array}(c), the repetitive pattern, large disparity, and occlusion converge in the same area (the zoom-in red rectangle), so the artifacts appear. 
This is a common failure mode for the methods which rely on texture matching cues to infer depth. In contrast, thanks to the depth-independent characteristic, our proposed method shows better performance under large disparities. Furthermore, it is noticed that the proposed method maintains better view consistency compared with other methods. Quantitative results on several Disney light fields are shown in the last four columns of Tab.\ref{tab:All Quantitative results}.

In addition, by comparing the results of our method in Tab.\ref{tab:All Quantitative results} with Tab.\ref{tab:EFS energy loss}, we find that the average EFS spectral energy loss has a significant influence on PSNR while having much less effect on SSIM. This demonstrates the robustness of the proposed reconstruction method, considering the fact that human eyes are more sensitive to changes of the structure information (SSIM) than individual pixel values (PSNR) \cite{ssim_paper}.

%LLFF \cite{mildenhall2019local} is based on MPI, and the MPI network sometimes assigns high opacity to incorrect layers in regions of ambiguous or repetitive texture. This can cause floating or blurred patches in the rendered output sequence, such as the far front part of the statue scene (Fig.\ref{fig:sota_camera_array}(c)). It is a common failure mode in methods that rely on texture matching cues to infer depth. Our proposed method is independent of depth cues, so the reconstruction results are better in large disparity areas, as shown in Fig.\ref{fig:sota_camera_array}(d). Furthermore, from the reconstructed EPIs, we can find that our method can preserve good consistency between views. The last four rows of Tab.\ref{tab:All Quantitative results} lists the average PSNR and SSIM values of different methods on four Disney light fields.

\begin{table*}[t!]
	\begin{center}
		\caption{Quantitative comparisons with SOTAs under different downsampling rates on both synthetic and real light fields.}
		\label{tab:All Quantitative results}
		%\begin{tabular}{lcccccccc}
		\begin{tabularx}{\textwidth}{@{} lCCCCCCCCC @{}}
			\hline\hline
			&& \multicolumn{2}{c}{Syn. LFs}& \multicolumn{2}{c}{ Real LFs \cite{guo2018dense}}& Bike \cite{kim2013scene} & Church \cite{kim2013scene} &Couch \cite{kim2013scene} & Statue \cite{kim2013scene} \\
			%\hline
			
			&& $10\times$ & $15\times$ &  $10\times$ & $15\times$ &  $5\times$ &$10\times$ & $10\times$ & $10\times$ \\
            \cmidrule(r){1-10}

             \multirow{3}{*}{Wu  \cite{wu2017light_epi}}& PSNR$\uparrow$ &35.48 &34.77 &36.14&	34.92&	 31.13 &32.64 &32.01 &32.63 \\

            &SSIM$\uparrow$ &0.851  &0.813	 &0.877 &	0.825	 &0.708 &0.721	 & 0.724& 0.698  \\ 
            &LPIPS$\downarrow$ &0.060 &0.093 &0.050 &0.078 &0.110 &0.080 &0.113 &0.098 \\
            \hline

          \multirow{3}{*}{LLFF  \cite{mildenhall2019local} }&PSNR$\uparrow$&37.29 &36.46    &39.74 &37.02 &35.51 &\textbf{38.85}&37.25 &38.53\\
          &SSIM$\uparrow$ &0.941 &0.922&\textbf{0.964} &0.925 &0.875 &0.962 &0.916 &0.956\\  

          &LPIPS$\downarrow$&0.059 &0.089 &\textbf{0.039} &0.079 &\textbf{0.082}&\textbf{0.051} &0.084 &0.049   \\

        \hline
       \multirow{3}{*}{Ours}&PSNR$\uparrow$& \textbf{39.36}& \textbf{37.54}& \textbf{40.18}& \textbf{37.74}& \textbf{36.77}& 37.95& \textbf{43.05}& \textbf{40.82}\\
        &SSIM$\uparrow$ & \textbf{0.963} &\textbf{0.952} &0.948 &\textbf{0.938} & \textbf{0.939} & \textbf{0.964} & \textbf{0.928} &\textbf{0.959}\\ 
        &LPIPS$\downarrow$& \textbf{0.056} &\textbf{0.088} &0.043 &\textbf{0.072} &0.085 &0.060 &\textbf{0.079} &\textbf{0.041}   \\

\hline\hline
		\end{tabularx}
	\end{center}
\end{table*}

\subsection{Multi-reference-view results}
As shown in Fig.\ref{fig:sota_syn_psnr_ssim} and Fig.\ref{fig:sota_real_psnr_ssim}, the reconstruction performance for some edge views by our method (only one  center reference) has been slightly below than that of the SOTAs. This is because the center reference view does not contain enough side-view information when the baseline is large. In this section, we use multi-reference views to build Multi-EFSs to provide more information for marginal view reconstruction.

\begin{figure}
\begin{center}
\centering
\includegraphics[width=\linewidth]{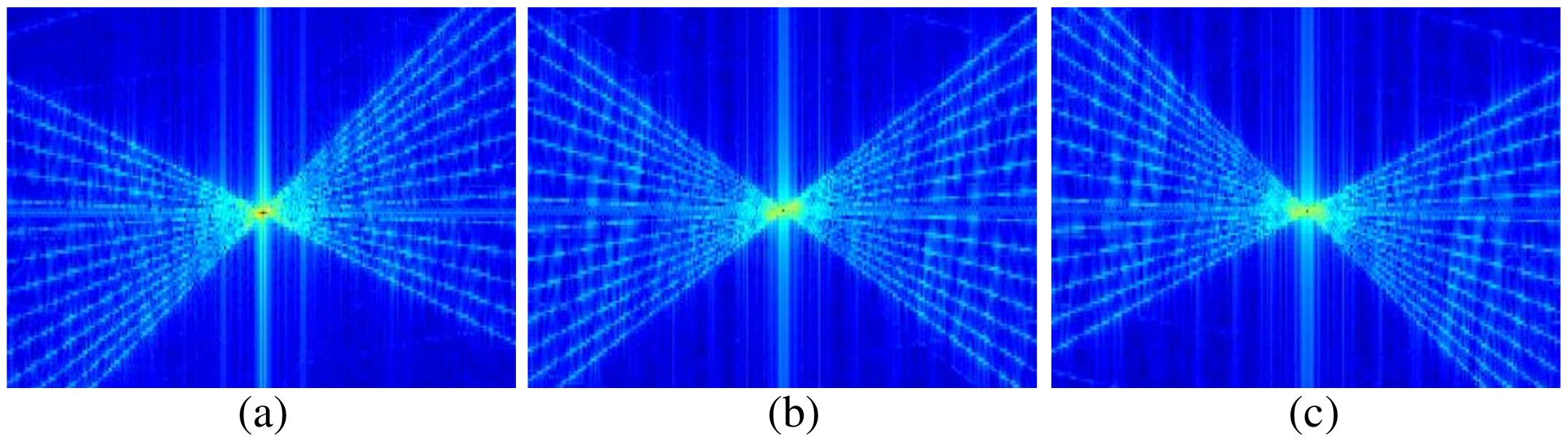}
\end{center}
\caption{Reconstructed EFSs obtained by different reference views under 15$\times$ downsampling. (a) EFS obtained by the 30th view ahead of the center view, (b) EFS obtained by the center view, (c) EFS obtained by the 30th view behind the center view. }
\label{fig:muti_input_efs}
\end{figure}

Fig.\ref{fig:muti_input_efs} show the reconstructed Multi-EFSs, which is built using the 30th view ahead of the center view ($u_{ref}=u_{cen}-30$), the center reference ($u_{ref}=u_{cen}$) and the 30th view behind the center view($u_{ref}=u_{cen}+30$). The bottom row of Fig.\ref{fig:sota_syn_psnr_ssim} and Fig.\ref{fig:sota_real_psnr_ssim} show the PSNR and SSIM measurements for each reconstructed view of Fig.\ref{fig:sota_syn} and Fig.\ref{fig:sota_realdata}. It can be seen that the Multi-EFSs significantly improve the overall performance of our EFS-based method, especially for edge views.

%\textcolor{red}{Fig.\ref{fig:muti_ref_psnr_ssim} shows the PSNR and SSIM measurements for each reconstructed view of Fig.\ref{fig:sota_syn} and Fig.\ref{fig:sota_realdata}, the Multi-EFSs improve the overall performance of our EFS-based method, especially for edge views. (partially repeated with Fig.19 and Fig.20, maybe combine these figures together?)}

%\begin{figure}
%\begin{center}
%\centering
%\includegraphics[width=\linewidth]{muti_ref_psnr_ssim.pdf}
%\end{center}
%\caption{Quantitative comparison of multi-reference-view result. PSNR and SSIM for each reconstructed view of Fig.\ref{fig:sota_syn} (the top row) and Fig.\ref{fig:sota_realdata} (the bottom row) under 15$\times$ downsampling.}
%\label{fig:muti_ref_psnr_ssim}
%\end{figure}

\begin{figure}
\begin{center}
\centering
\includegraphics[width=\linewidth]{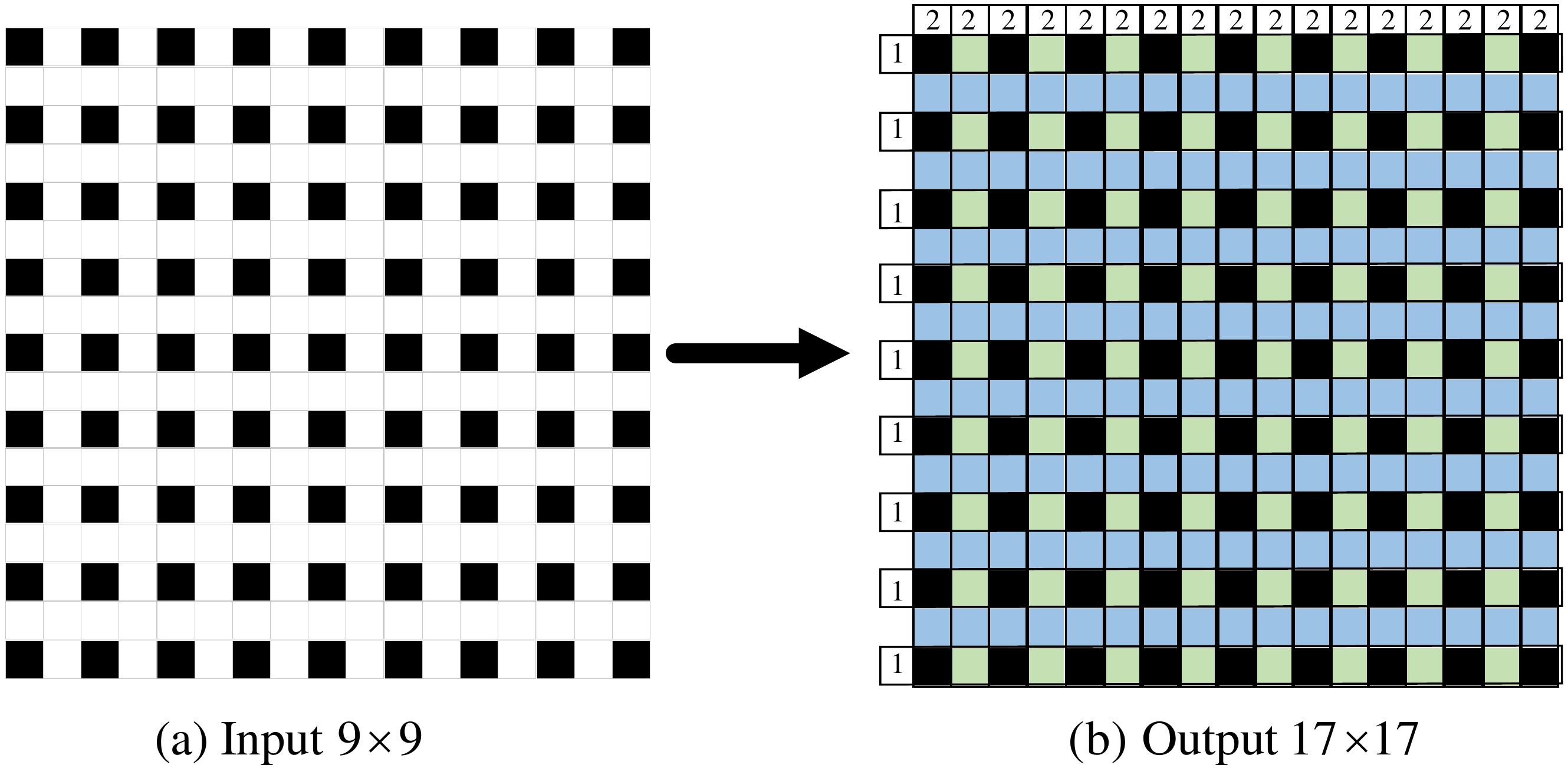}
\end{center}
\caption{Illustration of the sequential solution for full parallax light field reconstruction. (a) $9 \times 9$ input views. (b) $17 \times 17$ output views. }
\label{fig:solution_of_full_parallax}
\end{figure}

\begin{figure*}[h]
\begin{center}
\centering
\includegraphics[width=\linewidth]{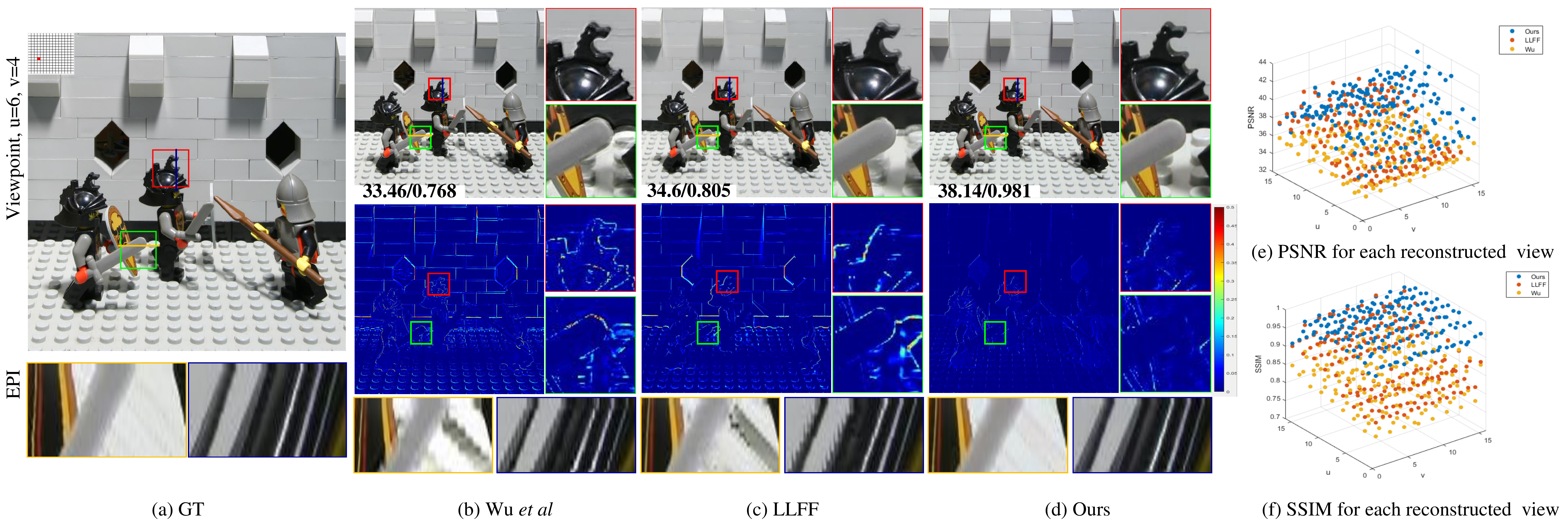}
\end{center}
\caption{Comparison of reconstruction results on the Lego dataset ($17 \times 17$ full parallax views reconstructed from $9 \times 9$ views). The results include one reconstructed view, the error map, and the EPI of the reconstructed light field by different methods. Several local areas are zoomed in for better visualization. From left to right: (a) GT, results by (b) Wu {\it {et al.}} \cite{wu2018light}, (c) LLFF \cite{mildenhall2019local}, and (d) our method, quantitative results in terms of (e) PSNR and (f) SSIM. For better comparison, the nearest neighbor algorithm is utilized to interpolate the EPI in this figure. }
\label{fig:legao_SOAT}
\end{figure*}

\subsection{Full parallax results}
\label{sec:Full parallax  result}
The sequential solution for full parallax is reconstructing the views containing vertical disparities after all the views containing horizontal disparities have been reconstructed, which is illustrated using an array of $17 \times 17$ full parallax reconstructed from the $9 \times 9$ input views (marked in black) in Fig.\ref{fig:solution_of_full_parallax}. As shown in Fig.\ref{fig:solution_of_full_parallax}(b), the views marked in green are firstly reconstructed in the horizontal parallax reconstruction step, and the views marked in blue are later reconstructed in the vertical parallax reconstruction step. We validate this solution on the Lego dataset\cite{stanford_lf_web} and show the qualitative and quantitative results in Fig.\ref{fig:legao_SOAT}. 

\begin{table}[t]
\setlength{\abovecaptionskip}{-0.1cm}  
\begin{center}
\caption{The average PSNR and SSIM on the Lego dataset.}

\label{tab:lego}
%\begin{tabular} {cccc} 
\begin{tabularx}{0.45\textwidth}{@{} CCCC @{}}
\hline\hline
      & Wu {\it {et al.}} \cite{wu2018light}& LLFF \cite{mildenhall2019local}& Ours\\
\hline
PSNR$\uparrow$&  36.29 &37.71 & \textbf{39.02}\\
SSIM$\uparrow$&  0.862 & 0.895 & \textbf{0.965}\\ \hline\hline
\end{tabularx}
	\end{center}
	\vspace{-0.5cm} 
\end{table}

Due to the lager disparity $([-9,7])$ of the Lego dataset, there exist obvious artifacts near boundary regions and discontinuous EPIs in both the results by Wu {\it {et al.}} \cite{wu2018light} and LLFF \cite{mildenhall2019local}. Additionally, LLFF \cite{mildenhall2019local} usually requires a large dataset for model training, 
%while our training datasets do not provide enough capacity for training this model. 
while our method is capable of learning the relationship between viewpoints and spectrum lines in the frequency domain from a relatively small training dataset. The experimental results show that our proposed method can generate clear edges and preserve cross-view consistency. Tab.\ref{tab:lego} shows the average PSNR and SSIM on the Lego dataset.
%More results for full parallax reconstruction (an array of $34 \times 34$ views reconstructed from $9 \times 9$ input views) are shown in the supplementary materials.

\subsection{Limitation}
\label{sec:Limitation}
%At present, in order to better analyse the relationship between EFS lines and the views, the proposed method only focuses on the disparity along one direction. A possible solution for full parallax is reconstructing the views containing vertical disparities after all views containing horizontal disparities have been reconstructed. 
At present, the energy loss of the spectrum still exists, which may lead to unevenly colored reconstructed views. A possible solution to deal with this issue could be the space-frequency combined method, or adding a visual channel to compensate for the energy loss in the frequency domain. In addition, although the proposed method outperforms the SOTA methods on both view reconstruction quality and cross-view consistency preservation, the shearing operation may cause a moderate decline in the reconstruction performance for the marginal views (Sec.\ref{sec:Comparison with SOTAs}). This could be addressed by introducing more reference views during the construction of the focal stack to provide more required information on the scene.

\section{Conclusions}
\label{sec:Conclusions}
We present a novel EFS representation for light fields, of which the advantages include: (1) it is depth-independent, (2) its structure remains basically unchanged under different angular sampling rates, and (3) each EFS-line corresponds to one view in the light field. Thanks to these characteristics, we further propose an EFS-based method for reconstructing a dense-view light field from an undersampled light field, by reconstructing EFS-lines in the spectrum domain without using scene depth or any spatial information. The proposed method exhibits superior performance under many challenging conditions, such as large disparities and complex occlusions, and maintains the cross-view consistency.

\ifCLASSOPTIONcaptionsoff
  \newpage
\fi

\bibliographystyle{IEEEtran}
\bibliography{main}

\end{document}